\documentclass[
pra,
aps,
reprint,
a4paper,
superscriptaddress,
longbibliography,
floatfix,
]{revtex4-2}

\usepackage{times}
\usepackage{graphicx}
\usepackage{float}
\usepackage{epsfig}
\usepackage{amsfonts}
\usepackage{amsmath}
\usepackage{amssymb}
\usepackage{color}
\usepackage{multirow}
\usepackage[colorlinks=true,linkcolor=blue,citecolor=blue,urlcolor=blue]{hyperref}
\usepackage[english]{babel}
\usepackage[utf8]{inputenc}
\usepackage[T1]{fontenc}
\usepackage{physics}
\usepackage{bm}
\usepackage{caption}
\usepackage{subcaption}
\usepackage{comment}

\begin{document}

\title{Stochastic optimization algorithms for quantum applications}

\author{J.~Gidi}
\email{jorgegidi@udec.cl}
\affiliation{Instituto Milenio de Investigaci\'on en \'Optica y Departamento de F\'isica, Facultad de Ciencias F\'isicas y Matem\'aticas, Universidad de Concepci\'on, Casilla 160-C, Concepci\'on, Chile}

\author{B.~Candia}
\affiliation{Instituto Milenio de Investigaci\'on en \'Optica y Departamento de F\'isica, Facultad de Ciencias F\'isicas y Matem\'aticas, Universidad de Concepci\'on, Casilla 160-C, Concepci\'on, Chile}

\author{A.~D. Muñoz-Moller}
\affiliation{Instituto Milenio de Investigaci\'on en \'Optica y Departamento de F\'isica, Facultad de Ciencias F\'isicas y Matem\'aticas, Universidad de Concepci\'on, Casilla 160-C, Concepci\'on, Chile}

\author{A.~Rojas}
\affiliation{Instituto Milenio de Investigaci\'on en \'Optica y Departamento de F\'isica, Facultad de Ciencias F\'isicas y Matem\'aticas, Universidad de Concepci\'on, Casilla 160-C, Concepci\'on, Chile}

\author{L.~Pereira}
\affiliation{Instituto de F\'{i}sica Fundamental IFF-CSIC, Calle Serrano 113b, Madrid 28006, Spain}

\author{M.~Muñoz}
\affiliation{Departamento de Ingenier\'ia Matem\'atica y Centro de Investigaci\'on en Ingenier\'ia Matem\'atica ($CI^2MA$), Facultad de Ciencias F\'isicas y Matem\'aticas, Universidad de Concepción, Casilla 160-C, Concepci\'on, Chile}

\author{L.~Zambrano}
\affiliation{ICFO - Institut de Ciencies Fotoniques, The Barcelona Institute of Science and Technology, 08860 Castelldefels, Barcelona, Spain}


\author{A.~Delgado}
\affiliation{Instituto Milenio de Investigaci\'on en \'Optica y Departamento de F\'isica, Facultad de Ciencias F\'isicas y Matem\'aticas, Universidad de Concepci\'on, Casilla 160-C, Concepci\'on, Chile}

\begin{abstract}
Hybrid classical quantum optimization methods have become an important tool for efficiently solving problems in the current generation of NISQ computers. These methods use an optimization algorithm executed in a classical computer, fed with values of the objective function obtained in a quantum processor. A proper choice of optimization algorithm is essential to achieve good performance. Here, we review the use of first-order, second-order, and quantum natural gradient stochastic optimization methods, which are defined in the field of real numbers, and propose new stochastic algorithms defined in the field of complex numbers. The performance of all methods is evaluated by means of their application to variational quantum eigensolver, quantum control of quantum states, and quantum state estimation. In general, complex number optimization algorithms perform best, with first-order complex algorithms consistently achieving the best performance, closely followed by complex quantum natural algorithms, which do not require expensive hyperparameters calibration. In particular, the scalar formulation of the complex quantum natural algorithm allows to achieve good performance with low classical computational cost.

\end{abstract}

\maketitle

\section{Introduction}

The current generation of quantum hardware has been described as noisy intermediate-scale quantum devices (NISQ)~\cite{Preskill2018}, characterized by noisy entangling gates, short coherence times, and large sampling errors. A promising approach to achieve quantum advantage in NISQ devices are hybrid quantum-classical optimization algorithms~\cite{Moll2018,McClean2016,RevModPhys.94.015004,2111.05176}. These evaluate an objective function through a parameterized quantum circuit in a quantum computer and feed the values of the objective function to a classical optimization algorithm running on a classical computer. Thus, hybrid optimization algorithms are used whenever the objective function can be evaluated more efficiently on a quantum computer than on a classical one. This is the case for applications to quantum chemistry~\cite{Lanyon2010,PhysRevX.8.031022,Nam2020}, quantum control~\cite{PhysRevA.91.052306,Lu2017,PhysRevLett.112.240503}, quantum simulation~\cite{Yuan2019,PhysRevLett.125.010501}, entanglement detection~\cite{Wang2021,2012.14311,2110.03709}, state estimation~\cite{PhysRevLett.113.190404, PhysRevLett.117.040402, CSPSA, Zambrano2020, Rambach2021}, quantum machine learning~\cite{Biamonte2017, Benedetti2019,PhysRevResearch.3.013063, Chen2020, Concha2012}, error correction~\cite{PhysRevApplied.15.034068}, graph theory~\cite{1411.4028,zhou2020quantum,Harrigan2021}, differential equations~\cite{PhysRevA.103.052425,PhysRevA.101.010301, PhysRevA.105.012433}, and finances~\cite{herman2022survey}.

The performance of hybrid quantum-classical algorithms is affected by the optimization landscape associated with the objective function and the choice of the optimization algorithm. For instance, it has been recently shown~\cite{McClean2018,Arrasmith2021} that a very general class of objective functions exhibit a barren plateau, that is a region in the optimization landscape where the objective function gradient vanishes and its standard deviation decreases exponentially with the number of qubits. In particular, this affects applications where random quantum circuits are used, such as, for instance, quantum machine learning~\cite{Benedetti2019}. Also recently, several studies~\cite{Sung2020, 2202.01389, Lavrijsen2020, 2111.13454} have been carried out to establish general guidelines to choose the optimization algorithm with the best performance, according to a predefined metric, for a certain class of problems. These consider methods such as Stochastic Gradient Descent~\cite{StochasticGradient1999}, Adaptive Gradient Algorithm~\cite{Duchi2011}, Root Mean Square Propagation~\cite{tieleman2012lecture}, Adam and variations~\cite{1412.6980,Dozat2016IncorporatingNM, 1904.09237}, Nelder-Mead~\cite{NelderMead}, Powell method~\cite{Powell}, and Newton Conjugate Gradient~\cite{Nash1984}, among many others~\cite{kraft1988software, Fletcher1964, Powell1994, Byrd1995, Lalee1998On, NumericalOptimization2006, Virtanen2020}.

In the growing list of optimization methods used in hybrid optimization, stochastic optimization algorithms~\cite{Kushner1978,Kushner1997,Albert2003,Spall2007,Bhatnagar2013} play an important role. State initialization, quantum gates, and measurements are noisy processes leading to noisy evaluation of the objective function. This intrinsically stochastic behavior of the objective function negates mathematical guarantees on the convergence of commonly used classical optimization methods~\cite{Lavrijsen2020}. However, certain stochastic optimization methods have convergence proofs that admit the presence of noise. In this scenario, a method that achieves good performance in various applications of hybrid optimization is the simultaneous perturbation stochastic approximation (SPSA) method~\cite{Spall1987ASA}. The main advantages of SPSA are its robustness to noise, ubiquitous in quantum mechanics, and that it can approximate the gradient of an objective function with only two measurements. In particular, this approximation does not require knowing the operational form of the objective function. SPSA has been successfully implemented in several experimental platforms and is one of the standards methods for training variational quantum eigensolvers (VQEs)~\cite{Kandala2017, Borzenkova2021, Hamamura2020, 2202.06979, PhysRevA.104.062426}, quantum neural networks (QNNs)~\cite{Benedetti2019_2,Mangini2020,Agliardi2022}, and quantum tomography \cite{PhysRevLett.113.190404,PhysRevLett.117.040402,PhysRevA.101.022317}. 

Given the success of stochastic optimization algorithms within quantum computing, efforts have been made to improve their performance in solving certain tasks. One proposal is second-order SPSA (2SPSA), which improves the convergence rate of SPSA by preconditioning the gradient with the inverse of a simultaneous perturbation estimate of the Hessian of the objective function~\cite{Spall2000,2012.06952}. This method is inspired by the deterministic Newton-Raphson algorithm and requires four evaluations of the objective function per iteration to estimate both gradient and Hessian. It has been shown that this method achieves a nearly optimal asymptotic error for well-conditioned problems. However, for a poorly conditioned Hessian, the error is several orders of magnitude larger~\cite{Zhu2002}. Another proposal focused on quantum computing is quantum natural gradient optimization~\cite{Stokes2020}. The SPSA algorithm explores the parameter space within a flat geometry, which can lead to an unfavorable update of parameters. In contrast, quantum natural gradient uses information about the geometry of the parametric quantum state to update the parameters appropriately. The Fubini-Study metric tensor represents this information. Natural gradient optimization provides several advantages over vanilla (or standard) methods, that is, methods in their unmodified form. This is because the natural gradient is invariant under re-parametrization~\cite{Amari1998} and approximately invariant under over-parametrization~\cite{Liang2019}. The version of SPSA based on the quantum natural gradient (QN-SPSA) uses a simultaneous perturbation estimate of the Fubiny-Study metric tensor~\cite{QN-SPSA}. This estimation requires four fidelity evaluations per iteration and the two function evaluations required to estimate the gradient. The fidelity evaluation can be performed efficiently using the swap-test~\cite{Buhrman2001}, among other alternatives~\cite{2103.09076}. This method is appropriate in contexts where the evaluation of the objective function is too expensive, for example, in estimating the fundamental energy of molecules~\cite{1907.13623,1701.08213,Hamamura2020,2202.06979}. However, similarly to 2SPSA, ill-conditioned metrics can reduce the performance of QN-SPSA \cite{1909.05074, PhysRevResearch.2.043246, PRXQuantum.2.030324}. 

Optimization methods can also be extended to work in the field of complex numbers by means of Wirtinger calculus \cite{Wirtinger1927}. Some examples are the complex Newton-Raphson algorithm \cite{Kreutz-Delgado2009} and the complex quantum natural gradient \cite{yao2021natural}. These methods optimize the objective function without resorting to the real and imaginary parts of complex variables. It has been argued in the literature that optimization methods formulated within the complex numbers could achieve better performance, which has been observed in a small set of examples~\cite{Zhang2015, Hirose2012, Smirnov2015}. This seems to be a more natural approach to optimization in quantum mechanics, where most functions have complex arguments. For example, continuous variable quantum computing employs displacement and squeezing operators, which depend on complex parameters \cite{Arrazola2019,PhysRevResearch.1.033063}. Recently, the complex simultaneous perturbation stochastic approximation (CSPSA) method~\cite{CSPSA} has been introduced. This is a generalization of SPSA that optimizes within the field of complex numbers. It has been shown that CSPSA can deliver better results in the estimation of pure states~\cite{CSPSA} and is robust against noise \cite{Rambach2021}. It has been applied to entanglement estimation~\cite{2110.03709}, quantum state discrimination~\cite{Concha2012}, and violation of the Claus-Horne-Shimony-Holt inequality~\cite{Cortes-Vega}.

Here, we present a comparative analysis of several stochastic optimization methods applied to real-valued functions of complex variables. We first review the basic principles of the SPSA algorithm. Subsequently, we review the 2SPSA and QN-SPSA algorithms using SPSA as a guideline. We also reviewed the CSPSA algorithm and developed two new optimization algorithms based on the CSPSA algorithm: 2CSPSA and QN-CSPSA. These are the complex field formulations of their real counterparts 2SPSA and QN-SPSA, respectively. We study the performance of the introduced methods by comparing their convergence rate as a function of the number of iterations with respect to SPSA, 2SPSA, and QN-SPSA. This comparison is carried out in three contemporary applications: variational quantum eigensolver, quantum control, and quantum state estimation. We use a variational quantum eigensolver to obtain the ground state energy of the Heisenberg Hamiltonian for a 10-qubit ring configuration, which is a ubiquitous and relatively simple model that describes the interactions within a chain of spins~\cite{Kandala2017}. We implement the GRadient Ascent Pulse Engineering (GRAPE) method~\cite{QC-GRAPE}, which is used to engineer quantum gates and states. This method approximates a control pulse by a sequence of constant-intensity pulses. The control parameters of this pulse are optimized to find the best implementation of a given gate or state, even in the presence of noise~\cite{PhysRevA.91.052306}. In particular, we apply GRAPE to the generation of 5-qubit pure states. Finally, in quantum state estimation, we implement Self-Guided Quantum Tomography (SGQT)~\cite{PhysRevLett.113.190404}, which is based on the minimization of the infidelity between an unknown state and a known parametrized state, to characterize 6-qubit pure states. Since the studied optimization methods are stochastic, we use numerical simulations and sampling to estimate the mean, variance (or standard deviation), median, and interquartile range of the relevant figures of merits. Measurements are simulated using a finite sample of various sizes.

Our comparative analysis shows that the first-order CSPSA algorithm consistently performs best in all three applications. In the case of variational quantum eigensolver, the performance of CSPSA is achieved at the expense of calibrating gain coefficients. Without the calibration, CSPSA performed poorly. To avoid the calibration, a good alternative is quantum natural algorithms, which achieve a performance close to the calibrated CSPSA at the expense of increasing the number of measurements and the classical computational cost. The latter can be avoided with the scalar version of quantum natural algorithms. On the other hand, for quantum state estimation and quantum control, the CSPSA algorithm performs the best without hyperparameter calibration.

While second-order algorithms do not provide an advantage, the quantum natural algorithms are competitive against first-order algorithms. As the number of qubits increases, we expect quantum natural algorithms to become more relevant. In this scenario, however, the cost of quantum natural algorithms increases. This can be partially mitigated using their scalar versions, which render the classical computational cost of quantum-natural algorithms feasible even for a very large number of qubits.

In general, complex-based optimization methods tend to outperform real-based optimization methods, although the difference in performance may be slight.

This article is organized as follows: in Section II, we review the stochastic optimization methods SPSA, 2SPSA, QN-SPSA, and CSPSA and formulate the methods 2CSPSA and QN-CSPSA. Also, we review and introduce modifications that may improve the performance of the methods. In Section III, we apply the previously developed optimization methods to variational quantum eigensolver, quantum control, and quantum state estimation. In section IV, we summarize our main results and conclusions.

\section{Stochastic optimization algorithms}

Let us consider the problem of optimizing a real function $f$ of $p$ complex variables, $f:\mathbb{C}^{p}\to\mathbb{R}$, that is, finding an argument $\bm z^{\star} \in \mathbb{C}^{p}$ such that $f(\bm z^{\star}$) is a local minimum of the function $f$. This problem can be solved by mapping the complex variables to the field of the real numbers through the relation $\bm z = \bm x + i\bm y$, in which case $f$ becomes $f(\bm\theta)$ with $\bm\theta = (\bm x, \bm y)^{T} \in \mathbb{R}^{2p}$. Then, one can use real variable optimization algorithms to find $\bm\theta^{\star} = (\bm x^{\star}, \bm y^{\star})^{T}$ such that $ f(\bm\theta^{\star})$ is a minimum of $f$, and retrieve the solution for the original complex variable problem as $\bm z^{\star} = \bm x^{\star } + i\bm y^{\star}$. It is possible, nevertheless, to solve the optimization problem using Wirtinger calculus~\cite{Wirtinger1927,Sorber2012}, which does not resort to mapping complex variables to real ones.

While both approaches are equivalent, the process of solving one or the other is not. It has been conjectured that a complex variable reformulation of real variable optimization algorithms may lead to increased performance \cite{Zhang2015,Hirose2012,Smirnov2015}, which has been observed when working on pure-state quantum tomography \cite{CSPSA}. Furthermore, for applications in quantum theory, which are natively stated in terms of complex variables, the transformation to real variables adds an extra step in the optimization process. For this reason, here we review some real variable optimization methods relevant to quantum applications and present their complex variable analogs.

A particularly suitable class of methods for optimizing multivariate functions in the presence of noisy measurements are the stochastic approximation methods \cite{Kushner1978,Kushner1997,Albert2003,Spall2007,Bhatnagar2013}. This family of methods originates from the Robbins-Monro algorithm \cite{Robbins-Monro} designed to find a root $\theta$ of a function $M(\bm x)$  given by the expectation of a random variable $Y(x)$. Here $M$ is unknown, just like the probability function of $Y$, and the Robbins-Monro algorithm gives an estimate of $\theta$ by making successive observations on $Y$. From the Robbins-Monro algorithm, it is possible to consider $M$ as a regression function \cite{Kiefer1952} and propose a scheme to estimate the maximum of $M$. Therefore, the use of stochastic approximations arises to deliver an algorithm that converges to an optimal value of a function $f$ using the Kiefer and Wolfowitz procedure when $M=\nabla f$.

A widely used family of stochastic approximation (SA) methods is based on the iterative rule 
\begin{align}
  \label{eq:first-order-iteration}
  \bm\theta_{k+1} &=  \bm\theta_{k} - a_{k}\bm g_{k}(\bm\theta_{k}),
\end{align}
where the descent step series $a_{k} = a/(k+A)^{s}$ is fixed by the externally selected gain parameters $a$, $A$, and $s$. The quantity $\bm g_{k}$ is a stochastic approximation of the gradient of the objective function at $\bm\theta_k$, which depends on the gain coefficient $b_k = b/k^t$, where $b$ and $t$ are externally fixed gain parameters.

In the following subsections, we review the SPSA algorithm and its extension to the second-order and quantum natural gradient algorithms, 2SPSA and QN-SPSA, respectively. Subsequently, we review the CSPSA algorithm for complex variables and develop two extensions to it; the second-order algorithm 2CSPSA and the quantum natural gradient algorithm QN-CSPSA. This work is conducted, in a similar way to the SPSA algorithm, by considering an iterative rule as Eq.~\eqref{eq:first-order-iteration} for the case of complex variables. Lastly, we present typical modifications to improve the performance of the optimization algorithms, namely blocking and resampling, and introduce two further variations: an alternative Hessian post-processing procedure and a scalar approximation to second-order and quantum natural algorithms that reduce their classical computational cost.

\subsection{Real-variable methods}

\subsubsection{SPSA}

The simultaneous perturbation stochastic approximation (SPSA) is a multivariate optimization method for real functions of real variables. While the SPSA denomination came later, the method was first presented by~\citet{Spall1987ASA} and corresponded to an improvement over the finite difference stochastic approximation (FDSA) from Kiefer and Wolfowitz~\cite{Kiefer1952}. Both the FDSA and SPSA algorithms optimize the function $f(\bm \theta)$ with $\bm\theta\in\mathbb{R}^{p}$ by following the recursive stochastic approximation rule Eq.~\eqref{eq:first-order-iteration}. However, the main feature of SPSA is that, instead of estimating each of the $p$ components of the gradient as a stochastic finite difference approximation, it defines the estimator $\bm g_{k}$ as
\begin{align}
  \label{eq:gradient-estimator}
  \bm g_{k}(\bm\theta) = \frac{f(\bm\theta + b_{k}\bm\Delta_{k}) - f(\bm\theta - b_{k}\bm\Delta_{k})}{2b_{k}}
  \begin{pmatrix}
    1/\Delta_{k,1} \\
    \vdots \\
    1/\Delta_{k, p}
  \end{pmatrix},
\end{align}
where $\bm\Delta_{k}$ is a random perturbation vector with $p$ components typically chosen from the set $\{\pm 1\}$ with uniform probability, and the finite-difference approximation step $b_{k}=b/k^{t}$ is controlled by the externally selected gain parameters $b$ and $t$. It is worth noting that while $\bm g_{k}(\bm\theta_{k})$ does not necessarily have the direction of the gradient at each iteration, it is an asymptotically unbiased estimator of the gradient, meaning that it converges at the statistical limit to the same solution as following the gradient. Furthermore, the Eq.~\eqref{eq:gradient-estimator} makes the SPSA algorithm especially suitable for high-dimensional problems since it always requires $2$ function evaluations per iteration, regardless of the number $p$ of variables, in contrast to the FDSA algorithm that requires $2p$ function evaluations per iteration.

An iteration of the SPSA algorithm is given by Eqs.~\eqref{eq:first-order-iteration} and \eqref{eq:gradient-estimator} and requires a total of $2$ objective function evaluations.

\subsubsection{2SPSA}

Since the iterative rule used in the SPSA algorithm is derived from a first-order gradient descent approximation, the rate of convergence of the algorithm could be accelerated using a second-order iterative rule coming from the Newton-Raphson method, given by
\begin{equation}
  \label{eq:Newton-Raphson}
  \bm\theta_{k+1} = \bm \theta_{k} - \eta \left[\mathcal{H}(\bm\theta_{k})\right]^{-1}\left(\frac{\partial f}{\partial\bm\theta}(\bm\theta_{k})\right)^{T},
\end{equation}
where $\eta \in \mathbb{R}^{+}$ is the learning rate and $\mathcal{H}$ is the Hessian of $f$. A stochastic approximation based on Eq.~(\ref{eq:Newton-Raphson}) is proposed by~\citet{Spall2000}, deriving the so-called adaptive or second-order SPSA (2SPSA) algorithm. The iterative rule now yields
\begin{align}
  \label{eq:second-order-iteration}
  \bm\theta_{k+1} &= \bm\theta_{k} - \overline{a}_{k}\overline{\mathcal{H}}_{k}^{-1}\bm g_{k}(\bm\theta_{k}),
\end{align}
where $\overline{a}_{k}=1/(k+A)^{s}$ no longer depends on $a$. The gradient estimator $\bm g_{k}$ is defined by Eq.~\eqref{eq:gradient-estimator}, as in the first-order case, and $\overline{\mathcal{H}}_{k }$ is a modified version of the simultaneous perturbations stochastic approximation of the Hessian matrix. In particular, we compute $\overline{\mathcal{H}}_{k}$ by~\cite{Spall2000}
\begin{subequations}
  \label{eq:hessian}
  \begin{align}
    \label{eq:hessian_symmetric}
    \mathcal{H}_{k}' &= \frac{\mathcal{H}_{k} + \left[\mathcal{H}_{k}\right]^{T}}{2},\\
    \label{eq:hessian_inertia}
    \mathcal{H}_{k}'' &= \frac{k}{k+1}\mathcal{H}_{k-1}'' + \frac{1}{k+1}\mathcal{H}_{k}',\\
    \label{eq:hessian_regularization}
    \overline{\mathcal{H}}_{k} &= \sqrt{\mathcal{H}_{k}''^{2}} + \varepsilon I,
  \end{align}
\end{subequations}
where, in execution order, Eq.~\eqref{eq:hessian_symmetric} ensures that the Hessian approximation is symmetric as the analytical Hessian, then Eq.~\eqref{eq:hessian_inertia} stabilizes the estimator by introducing inertia from previous iterations, starting from an identity at the zeroth iteration, $\mathcal{H}_{0}''=I$, and finally, Eq.~\eqref{eq:hessian_regularization} with ${0 < \varepsilon \ll 1}$ guarantees positive-definiteness.

A one-sided simultaneous perturbation stochastic approximation to the Hessian matrix is taken as
\begin{align}
  \label{eq:hessian_approx}
  \mathcal{H}_{k}(\bm\theta) &= \frac{\bm g_{k}(\bm\theta +\tilde{b}_{k}\tilde{\bm\Delta}_{k}) - \bm g_{k}(\bm\theta)}{\tilde{b}_{k}}
                               \begin{pmatrix}
                                 1/\tilde\Delta_{k, 1} \\
                                 \vdots \\
                                 1/\tilde\Delta_{k, p}
                               \end{pmatrix}^{T},
\end{align}
which allows reusing the function evaluations from the centered gradient estimator. By inserting the definition of the gradient approximation Eq.~(\ref{eq:gradient-estimator}), then Eq.~\eqref{eq:hessian_approx} can be rewritten by components as
\begin{align}
\label{eq:hessian_approx_components}
  \left[\mathcal{H}_{k}\right]_{ij} &= \frac{\delta^{2}f_{k}(\bm\theta)}{2b_{k}\tilde{b}_{k}\Delta_{k,i}\tilde\Delta_{k,j}},
\end{align}
where
\begin{align}
  \label{eq:d2f}
  \delta^{2}f_{k}(\bm\theta) &= f(\bm\theta + b_{k}\bm\Delta_{k} + \tilde{b}_{k}\tilde{\bm\Delta}_{k}) - f(\bm\theta + b_{k}\bm\Delta_{k})\nonumber\\
                             &\quad - f(\bm\theta - b_{k}\bm\Delta_{k} + \tilde{b}_{k}\tilde{\bm\Delta}_{k})  + f(\bm\theta-b_{k}\bm{\Delta}_{k}),
\end{align}
$\tilde{b}_{k}=\tilde b / k^{t}$ is a gain series similar in nature to $b_{k}$, and $\tilde{\bm\Delta}_{k}$ is a random vector formed by $p$ components uniformly generated from the set $\{\pm 1\}$ analogous to $\bm\Delta_{k}$.

Thereby, an iteration of this method is given by Eqs.~\eqref{eq:d2f},~\eqref{eq:hessian_approx_components},~\eqref{eq:hessian},~\eqref{eq:gradient-estimator}, and~\eqref{eq:second-order-iteration} and requires a total of $4$ objective function evaluations.

\subsubsection{QN-SPSA}

The Gradient Descent method reaches a local minimum by moving, at each iteration, along the direction of the steepest descent of the objective function in the Euclidian parameter space, $-(\partial f /\partial\bm\theta)^T$, limiting the magnitude of the
update step, $\Delta\bm\theta$. The steepest descent rule can be obtained by choosing the increment as
\begin{align}
  \label{eq:gradient_descent_problem}
  \Delta\bm\theta = \underset{\Delta\bm\theta\in\mathbb{R}^{2p}}{\arg\,\min} \left\lbrace \Big\langle
  {\left(\frac{\partial f}{\partial\bm\theta}\right)}^{T}, \Delta\bm\theta \Big\rangle
  + \frac{1}{2\eta}\big\|\Delta\bm\theta\big\|^{2}_{2} \right\rbrace,
\end{align}
where $\eta\in\mathbb{R}^{+}$ is the learning rate, $\langle \bm\theta,\bm\theta' \rangle = \bm\theta^{T}\bm\theta'$ is the inner product for two vectors $\bm\theta$ and $\bm\theta'$, respectively, and ${\|\cdot\|_{2}=\sqrt{\langle\cdot,\cdot\rangle}}$ is the $l^{2}$ norm. Differentiating the argument at the right-hand side of Eq.~\eqref{eq:gradient_descent_problem} with respect to $\Delta\bm\theta$ and setting it to $\bm 0$, provides the well-known gradient descent step
\begin{align}
  \Delta\bm\theta = -\eta\left( \frac{\partial f}{\partial\bm\theta} \right)^{T}.
\end{align}
This result is based on the $l^2$ geometry, where a shift in any direction in the parameter space is equally weighted. However, the objective function may not be equally sensitive to changes in different parameters and, therefore, a more adequate notion of distance would measure the step length $\Delta\bm\theta$ by weighting the changes on each parameter. This is addressed by a method called natural gradient descent~\cite{Amari}, which endows the parameter space with a suitable metric $\mathcal{G}$ that induces the norm $\|\cdot\|_{\mathcal{G}}=\sqrt{\langle\cdot,\mathcal{G}\,\cdot\rangle}$. Then, the increment is stated as
\begin{align}
  \label{eq:natural_gradient_problem}
  \Delta\bm\theta = \underset{\Delta\bm\theta\in\mathbb{R}^{2p}}{\arg\,\min} \left\lbrace \Big\langle
  {\left(\frac{\partial f}{\partial\bm\theta}\right)}^{T}, \Delta\bm\theta \Big\rangle
  + \frac{1}{2\eta}\big\|\Delta\bm\theta\big\|^{2}_{\mathcal{G}} \right\rbrace,
\end{align}
which leads to the natural gradient descent rule
\begin{align}
  \label{eq:natural-gradient-descent}
  \bm\theta_{k+1} = \bm\theta_{k} - \eta \left[\mathcal{G}(\bm\theta_{k})\right]^{-1}\left(\frac{\partial f}{\partial \bm\theta}(\bm\theta_{k})\right)^{T}.
\end{align}

The quantum natural method, which takes $\mathcal{G}$ as the Fubini-Study metric tensor, is particularly useful for improving convergence rates for optimization problems in quantum applications~\cite{Stokes2020}. The Fubini-Study metric tensor is proportional to the Quantum Fisher information matrix, so its calculation can be very expensive when many variables are involved. This problem was addressed by Gacon~\textit{et al.}~\cite{QN-SPSA} by taking advantage of the similarity between the Eqs.~\eqref{eq:Newton-Raphson} and \eqref{eq:natural-gradient-descent}, along with the possibility of writing the Fubini-Study metric tensor as
\begin{align}
  \label{eq:QFI}
  \mathcal{G}(\bm\theta) = \left.-\frac{1}{2} \left[ \frac{\partial}{\partial \bm\theta} \left(\frac{\partial F(\bm\theta', \bm\theta)}{\partial\bm\theta}\right)^{T} \right]\right|_{\bm\theta' = \bm\theta},
\end{align}
where $F(\bm\theta', \bm\theta)$ is the fidelity between two quantum states parameterized with the variables $\bm\theta'$ and $\bm\theta$, respectively. In particular, the Fubini-Study metric tensor was approximated according to Eq.~\eqref{eq:QFI} using the stochastic approximation of the Hessian employed by the 2SPSA algorithm.
In this manner, they proposed the quantum natural SPSA (QN-SPSA) algorithm, which avoids the curse of dimensionality.

In order to reuse the equations already presented for the 2SPSA method, we will abuse notation and denote $\mathcal{H}$ the Hessian estimate of the Fubini-Study metric, yielding
\begin{align}
  \label{eq:QN_hessian_approx}
  \left[\mathcal{H}_{k}\right]_{ij} = -\frac{\delta^{2}F_{k}(\bm\theta_{k})}{4b_{k}\tilde{b}_{k}\Delta_{k, i}\tilde\Delta_{k, j}},
\end{align}
where
\begin{align}
  \label{eq:d2F}
  \delta^{2}F_{k}(\bm\theta) &= F(\bm\theta, \bm\theta + b_{k}\bm\Delta_{k} + \tilde{b}_{k}\tilde{\bm\Delta}_{k}) \nonumber\\
                             &\quad - F(\bm\theta, \bm\theta + b_{k}\bm\Delta_{k})\nonumber\\
                             &\quad - F(\bm\theta, \bm\theta - b_{k}\bm\Delta_{k} + \tilde{b}_{k}\tilde{\bm\Delta}_{k})\nonumber\\
                             &\quad + F(\bm\theta, \bm\theta - b_{k}\bm\Delta_{k}),
\end{align}
and $\bm\Delta$ and $\tilde{\bm\Delta}$ are two vectors of $p$ components randomly sampled from the set $\{\pm 1\}$.

Following the same logic as in the 2SPSA algorithm, the simultaneous perturbation stochastic approximation of the Hessian Eq.~\eqref{eq:QN_hessian_approx} must be conditioned by the procedure on the system of Eqs.~\eqref{eq:hessian}. Let us note that while we are using the 2SPSA discretization scheme and update rule, this is a first-order method, as the conditioner $\mathcal{H}_{k}$ comes not from a second-order expansion on the target function but only from a different metric in the parameter space.

Requiring only two measurements of the objective function makes this algorithm especially suitable for problems where the metric tensor can be efficiently approximated. That is when evaluating the fidelity $F$ between two known pure quantum states requires marginal resources compared to the potentially expensive target function $f$.

An iteration of the QN-SPSA method is given by Eqs.~\eqref{eq:d2F}, \eqref{eq:QN_hessian_approx},~\eqref{eq:hessian}, \eqref{eq:gradient-estimator}, and \eqref{eq:second-order-iteration} and requires a total of $2$ objective function evaluations and $4$ fidelity evaluations.

\subsection{Complex-variable methods}

Now we formulate the problem of optimizing real-valued functions of complex variables. In the case of quantum mechanics, most of the functions that interest us depend on complex variables and their complex conjugates. Consequently, these functions do not satisfy the Cauchy-Riemann conditions and lack a Taylor series expansion. This can be solved by resorting to the real and imaginary parts of the complex variables. Wirtinger calculus~\cite{Wirtinger1927}, however, allows us to define a derivative, the Wirtinger derivative, that exists even for non-holomorphic functions. We consider a function $f: \bm\mu \in \mathbb{C}^{2p}\to \mathbb{R}$ with ${\bm\mu = (\bm z, \bm{z^{*}})^{T}}$, which can be expressed in a power series for a complex increment $\Delta \bm\mu = (\Delta\bm z, \Delta\bm z^{*})^{T}$,
\begin{align}
  \label{eq:mu-expansion}
  f(\bm\mu + \Delta\bm\mu) &= f(\bm\mu) + \frac{\partial f}{\partial\bm\mu}\Delta\bm\mu + \frac{1}{2}\Delta\bm\mu^{\dagger}
                             \mathcal{H}
                             \Delta\bm\mu + \dots,
\end{align}
where
\begin{align}
    \mathcal{H} = \frac{\partial}{\partial\bm\mu} \left( \frac{\partial f}{\partial\bm\mu}\right)^{\dagger}
\end{align}
is the complex Hessian of the function $f$~\cite{Kreutz-Delgado2009}, the symbol ($\dagger$) denotes the conjugate transpose, and differentiation with respect to $\bm\mu$ is defined by
\begin{align}
  \frac{\partial f}{\partial\bm\mu} = \left(
  \frac{\partial f}{\partial\bm z}~,~
  \frac{\partial f}{\partial\bm z^{*}}
  \right),
\end{align}
where the complex variables $\bm z$ and $\bm z^*$ are considered to be independent. Let us note that the inner product between any column two vectors ${\bm\mu=(\bm z\quad \bm z^{*})^{T}}$ and ${\bm\mu'=(\bm z'\quad \bm z'^{*})^{T}}$, with $\bm z, \bm z' \in \mathbb{C}^{p}$, is a real number,
\begin{align}
  \bm\mu^{\dagger}\bm\mu' = (\bm z^{*}\quad \bm z)
  \begin{pmatrix}
    \bm z' \\
    \bm z'^{*}
  \end{pmatrix}
  = 2\Re\{\bm z^{\dagger} \bm z'\}.
\end{align}

\subsubsection{CSPSA}

Performing a first-order approximation on $|\Delta\bm\mu|$ from Eq.~\eqref{eq:mu-expansion}, that is,
\begin{align}
    f(\bm\mu + \Delta\bm\mu) - f(\bm\mu) \approx \frac{\partial f}{\partial\bm\mu}\Delta\bm\mu,
\end{align}
we obtain that the largest decrease of the function $f$ is achieved by a perturbation $\Delta\bm\mu$ in the direction of $-(\partial f/\partial\bm\mu)^\dagger$. This provides the complex equivalent to the gradient descent update rule, which is given by the expression
\begin{align}
  \label{eq:complex_gradient_descent}
  \bm\mu_{k+1} = \bm\mu_{k} - \eta \left(\frac{\partial f}{\partial \bm\mu}\right)^{\dagger},
\end{align}
where $\eta\in\mathbb{R}^{+}$ is the learning rate. The above equation yields a stochastic approximation \cite{CSPSA} used to introduce the CSPSA algorithm given by the iterative rule
\begin{equation}
  \label{eq:complex-iteration}
  \bm z_{k+1} = \bm z_{k} - a_{k}\bm g_{k}(\bm z_{k}),
\end{equation}
where $a_{k} = a/(k+A)^{s}$. The gradient estimator is now given by
\begin{align}
  \label{eq:complex-gradient-estimator}
  \bm g_{k}(\bm z) = \frac{f(\bm z + b_{k}\bm\Delta_{k}) - f(\bm z - b_{k}\bm\Delta_{k})}{2b_{k}}
  \begin{pmatrix}
    1/\Delta_{k,1}^{*} \\
    \vdots \\
    1/\Delta_{k,p}^{*}
  \end{pmatrix},
\end{align}
where $b_{k}=b/k^{t}$ and $\bm\Delta_{k}$ is a random vector with $p$ components uniformly generated from the set $\{\pm 1, \pm i\}$, with $i$ the imaginary unit. To keep the notation simple, we have omitted the dependency of $\bm g_{k}$ on $\bm z^*$. Consequently, we write $\bm g_{k}(\bm z, \bm z^*)$ as $\bm g_{k}(\bm z)$ and similarly for other functions.

An iteration of the CSPSA method is given by Eqs.~\eqref{eq:complex-gradient-estimator} and \eqref{eq:complex-iteration} and requires a total of $2$ objective function evaluations.

\subsubsection{2CSPSA}

To obtain a second-order iterative rule, we add up to second-order terms on $|\Delta\bm\mu|$ from expansion Eq.~\eqref{eq:mu-expansion} and consider the problem of finding the perturbation $\Delta\bm\mu$ that minimizes $f(\bm\mu + \Delta\bm\mu)$. This is done by taking $\partial f(\bm\mu + \Delta\bm\mu)/\partial\Delta\bm\mu = 0$, which reduces to the equation
\begin{align}
  \label{eq:second-order-optimal-perturbation}
  \left[ \frac{\partial}{\partial\bm\mu} \left( \frac{\partial f}{\partial\bm\mu}\right)^{\dagger} \right]
  {\Delta\bm\mu} = -\left(\frac{\partial f}{\partial\bm\mu}\right)^{\dagger}.
\end{align}
This can be rewritten in terms of $\bm z$ and $\bm z^{*}$ as
\begin{align}
  \label{eq:second-order-update-z}
  \begin{pmatrix}
    \mathcal{H}_{zz}& \mathcal{H}_{zz^{*}} \\
    \mathcal{H}_{z^{*}z}& \mathcal{H}_{z^{*}z^{*}} \\
  \end{pmatrix}
  \begin{pmatrix}
    {\Delta\bm z} \\
    {\Delta\bm z^{*}}
  \end{pmatrix}
  = -\begin{pmatrix}
    [\partial f / \partial \bm z]^{\dagger} \\
    [\partial f / \partial \bm z^{*}]^{\dagger} \\
  \end{pmatrix},
\end{align}
where the elements of the block matrix are
\begin{align}
  \mathcal{H}_{zz} &= \frac{\partial}{\partial\bm z}\left(\frac{\partial f}{\partial\bm z}\right)^{\dagger},\\
  \mathcal{H}_{zz^{*}} &= \frac{\partial}{\partial\bm z}\left(\frac{\partial f}{\partial\bm z^{*}}\right)^{\dagger},\\
  \mathcal{H}_{z^{*}z} &= \mathcal{H}_{zz^{*}}^{\dagger},\text{ and} \\
  \mathcal{H}_{z^{*}z^{*}} &= \mathcal{H}_{zz}^{*}.
\end{align}
The system of Eqs.~\eqref{eq:second-order-update-z} has the solution
\begin{align}
  \label{eq:second-order-newton}
  {\Delta\bm z} &= \left( \mathcal{H}_{zz} - \mathcal{H}_{z^{*} z}\mathcal{H}_{z^{*}z^{*}}^{-1}\mathcal{H}_{zz^{*}} \right)^{-1} \nonumber \\
                        & \qquad\times \left\lbrace \mathcal{H}_{z^{*}z} \mathcal{H}_{z^{*}z^{*}}^{-1} \left(\frac{\partial f}{\partial \bm z^{*}}\right)^{\dagger} - \left(\frac{\partial f}{\partial \bm z}\right)^{\dagger} \right\rbrace,
\end{align}
which is the update step corresponding to a Newton algorithm \cite{Kreutz-Delgado2009}. While this solution requires a large number of operations, it is customary to use a block-diagonal approximation, $\mathcal{H}_{zz^{*}}\approx0$, yielding a pseudo-Newton method~\cite{GuangrongYan2000} with
\begin{align}
  \label{eq:second-order-pseudo-newton}
  {\Delta\bm z} = -\mathcal{H}_{zz}^{-1}\left(\frac{\partial f}{\partial\bm z}\right)^{\dagger},
\end{align}
which also has the advantage of being operationally independent of $\bm z^{*}$ in practice. 

Analog to the 2SPSA method, in the stochastic approximation, we take the descent direction given by Eq.~\eqref{eq:second-order-pseudo-newton}. Thereby, we define the 2CSPSA algorithm by means of the update rule
\begin{align}
  \label{eq:complex-second-order-iteration}
  \bm z_{k+1} &= \bm z_{k} - \overline{a}_{k}\left[\overline{\mathcal{H}}_{k}(\bm z_{k}) \right]^{-1} \bm g_{k}(\bm z_{k}),
\end{align}
where $\overline{a}_{k}=1/(k+A)^{s}$, $\bm g_{k}(\bm z)$ is given by Eq.~\eqref{eq:complex-gradient-estimator}, and $\overline{\mathcal{H}}_{k}$ is a modified version of the simultaneous perturbation stochastic approximation for the partial complex Hessian $\mathcal{H}_{zz}$ at the $k-$th iteration. Similar to the system of Eqs.~\eqref{eq:hessian} for the real-variable case, $\overline{\mathcal{H}}_{k}$ is computed through the sequence
\begin{subequations}
  \label{eq:complex_hessian}
  \begin{align}
    \label{eq:complex_hessian_hermitian}
    \mathcal{H}_{k}' &= \frac{\mathcal{H}_{k} + \left[\mathcal{H}_{k}\right]^{\dagger}}{2},\\
    \label{eq:complex_hessian_inertia}
    \mathcal{H}_{k}'' &= \frac{k}{k+1}\mathcal{H}_{k-1}'' + \frac{1}{k+1}\mathcal{H}_{k}',\\
    \label{eq:complex_hessian_regularization}
    \overline{\mathcal{H}}_{k} &= \sqrt{\mathcal{H}_{k}''^{2}} + \varepsilon I,
  \end{align}
\end{subequations}
where, in execution order, Eq.~\eqref{eq:complex_hessian_hermitian} makes the Hessian approximation hermitian as the exact Hessian, then Eq.~\eqref{eq:complex_hessian_inertia} stabilizes the estimator by introducing inertia from previous iterations, starting from an identity at the zeroth iteration, that is, $\mathcal{H}_{0}''=I$, and finally Eq.~\eqref{eq:complex_hessian_regularization} with ${0<\varepsilon \ll 1}$ ensures positive-definiteness. Note that the regularization Eq.~\eqref{eq:complex_hessian_regularization} is still valid in the complex-variable case since its input, $\mathcal{H}_{k}''$, has real eigenvalues due to the previous hermitization Eq.~\eqref{eq:complex_hessian_hermitian}.

In this case, the components of the simultaneous perturbation stochastic approximation of the partial complex Hessian $\mathcal{H}_{zz}$ are given by
\begin{align}
  \label{eq:complex_hessian_approx}
  \left[\mathcal{H}_{k}(\bm z)\right]_{ij} &= \frac{\delta^{2}f_{k}(\bm z)}{2b_{k}\tilde{b}_{k}\Delta_{k,i}^{*}\tilde{\Delta}_{k,j}},
\end{align}
where
\begin{align}
  \label{eq:complex_d2f}
  \delta^{2}f_{k}(\bm z) &= f(\bm z + b_{k}\bm\Delta_{k} + \tilde{b}_{k}\tilde{\bm\Delta}_{k}) - f(\bm z + b_{k}\bm\Delta_{k})\nonumber\\
                         &\quad - f(\bm z - b_{k}\bm\Delta_{k} + \tilde{b}_{k}\tilde{\bm\Delta}_{k}) + f(\bm z - b_{k}\bm\Delta_{k}),
\end{align}
and $\bm\Delta$ and $\tilde{\bm{\Delta}}$ are two random vectors, each composed by $p$ elements uniformly generated from the set $\{\pm 1, \pm i\}$.

The 2CSPSA method requires the inversion and regularization of a $p\times p$ hermitian complex matrix. In contrast, the analog 2SPSA optimization of an equivalent problem would require the inversion and regularization of a $2p\times 2p$ symmetric real matrix.

An iteration of this method is given by Eqs.~\eqref{eq:complex_d2f}, \eqref{eq:complex_hessian_approx}, \eqref{eq:complex_hessian}, \eqref{eq:complex-gradient-estimator}, and \eqref{eq:complex-second-order-iteration} and requires a total of $4$ objective function evaluations.

\subsubsection{QN-CSPSA}

The natural gradient method was adapted \cite{yao2021natural} for a complex parameter space by posing the usual natural gradient update rule Eq.~\eqref{eq:natural-gradient-descent} with the relevant metric $\mathcal{G}$ and using an invertible linear transformation $W$ to move back and forth between the real and complex parametrizations such that
\begin{align}
  W
  \begin{pmatrix}
    \bm x \\
    \bm y
  \end{pmatrix}
  =
  \begin{pmatrix}
    \bm x + i\bm y\\
    \bm x - i\bm y
  \end{pmatrix}
  :=
  \begin{pmatrix}
    \bm z\\
    \bm z^{*}
  \end{pmatrix},
\end{align}
where $\bm x, \bm y \in \mathbb{R}^{p}$. However, continuously moving between parameterizations is undesirable, and therefore here we present a natively complex implementation of the natural gradient method for quantum applications, which proceeds analogously to the QN-SPSA method.

The complex gradient descent rule Eq.~\eqref{eq:complex_gradient_descent} can be obtained as a solution to the optimization problem
\begin{align}
  \label{eq:complex_gradient_problem}
  \Delta\bm\mu = \underset{\Delta\bm\mu\in\mathbb{C}^{2p}}{\arg\,\min} \left\lbrace \Big\langle
  \left(\frac{\partial f}{\partial\bm\mu}\right)^{\dagger}, \Delta\bm\mu \Big\rangle
  + \frac{1}{2\eta}\big\|\Delta\bm\mu\big\|^{2}_{2} \right\rbrace,
\end{align}
where $\eta\in\mathbb{R}^{+}$ is the learning rate, $\langle \bm\mu,\bm\mu' \rangle = \bm\mu^{\dagger}\bm\mu'$ is the inner product for two complex vectors $\bm\mu$ and $\bm\mu'$, respectively, and $\|\cdot\|_{2}=\sqrt{\langle\cdot,\cdot\rangle}$ is the $l^{2}$ norm. As in the real case, to require the parameter update to remain small in the space endowed with metric $\mathcal{G}$, the $l^{2}$ norm is replaced in Eq.~\eqref{eq:complex_gradient_problem} by $\|\cdot\|_{\mathcal{G}} = \sqrt{\langle\cdot,\mathcal{G}\,\cdot\rangle}$. This leads to the optimization problem
\begin{align}
  \label{eq:complex_natural_gradient_problem}
  \Delta\bm\mu = \underset{\Delta\bm\mu\in\mathbb{C}^{2p}}{\arg\,\min} \left\lbrace \Big\langle
  \left(\frac{\partial f}{\partial\bm\mu}\right)^{\dagger}, \Delta\bm\mu \Big\rangle
  + \frac{1}{2\eta}\big\|\Delta\bm\mu\big\|^{2}_{\mathcal{G}} \right\rbrace,
\end{align}
which has the solution
\begin{align}
  \label{eq:complex_natural_gradient_step}
  \Delta\bm\mu = -\eta\mathcal{G}^{-1}\left(\frac{\partial f}{\partial\bm\mu}\right)^{\dagger},
\end{align}
where $\mathcal{G}$ is an hermitian matrix.

In the case that the optimization space is the set of pure quantum states, the metric $\mathcal{G}$ can be chosen proportional to the Quantum Fisher complex information matrix \cite{Munoz2022}, that is,
\begin{align}
  \label{eq:complex-Hessian-FS-metric}
  \mathcal{G} = -\frac{1}{2}\left.\left[ \frac{\partial}{\partial\bm\mu}\left( \frac{\partial F(\bm\mu', \bm\mu)}{\partial \bm\mu} \right)^{\dagger}\right]\right|_{\bm\mu' = \bm\mu},
\end{align}
where $F(\bm\mu', \bm\mu)$ is the fidelity between two states parameterized with variables $\bm\mu'$ and $\bm\mu$, respectively. 

Considering, as in the 2CSPSA case, a block-diagonal approximation of $\mathcal{G}$, the first row of Eq.~\eqref{eq:complex_natural_gradient_step} yields
\begin{align}
  \label{eq:complex_natural_gradient_update}
  \Delta\bm z = -\eta\mathcal{G}_{zz}^{-1}\left(\frac{\partial f}{\partial\bm z}\right)^{\dagger},
\end{align}
where $\mathcal{G}_{zz}$ is the top left block of $\mathcal{G}$.

Given the Hessian form of $\mathcal{G}_{zz}$ and considering the similarity of Eqs.~\eqref{eq:complex_natural_gradient_update} and~\eqref{eq:second-order-pseudo-newton}, we can borrow the discretization scheme from 2CSPSA to approximate $\mathcal{G}_{zz}$.
Denoting $\mathcal{H}_k$ as the simultaneous perturbation stochastic approximation of $\mathcal{G}_{zz}$ at iteration $k$, allows us to reuse the equations already presented for 2CSPSA giving
\begin{align}
  \label{eq:complex_QN_hessian_approx}
  \left[\mathcal{H}_{k}\right]_{ij} = -\frac{\delta^{2}F_{k}(\bm z_{k})}{4b_{k}\tilde{b}_{k}\Delta_{k,i}^{*}\tilde\Delta_{k,j}},
\end{align}
with
\begin{align}
  \label{eq:complex_d2F}
  \delta^{2}F_{k}(\bm z) &= F(\bm z, \bm z + b_{k}\bm\Delta_{k} + \tilde{b}_{k}\tilde{\bm\Delta}_{k}) \nonumber\\
                         &\quad - F(\bm z, \bm z + b_{k}\bm\Delta_{k})\nonumber\\
                         &\quad - F(\bm z, \bm z - b_{k}\bm\Delta_{k} + \tilde{b}_{k}\tilde{\bm\Delta}_{k})\nonumber\\
                         &\quad + F(\bm z, \bm z - b_{k}\bm\Delta_{k}),
\end{align}
and conditioning as in the system of Eqs.~\eqref{eq:complex_hessian}. As before, $\bm\Delta$ and $\tilde{\bm{\Delta}}$ are two random vectors, each composed by $p$ elements uniformly generated from the set $\{\pm 1, \pm i\}$.

An iteration of this method, which we call quantum natural CSPSA (QN-CSPSA), is given by Eqs.~\eqref{eq:complex_d2F}, \eqref{eq:complex_QN_hessian_approx}, \eqref{eq:complex_hessian}, \eqref{eq:complex-gradient-estimator}, and \eqref{eq:complex-second-order-iteration} and requires a total of $2$ objective function evaluations and $4$ fidelity evaluations.

\subsection{Method Improvements}
\label{sec:improv}

In the previous sections, optimization methods were presented in their vanilla form. It is possible, however, to introduce further modifications that can improve their convergence properties. In particular, we will address two typical modifications, blocking and resampling, and two extra variations we propose: an alternative Hessian post-processing procedure and a scalar approximation for the preconditioned methods.

\subsubsection{Blocking}

This technique consists of blocking the progression of the method if the updated parameters $\bm{z}_{k+1}$ fail to fulfill a given criterion. Conventionally, the updated variable is required to improve the value of the objective function with respect to the previous iteration plus some fixed non-negative tolerance,
\begin{equation}
  \label{eq:blocking}
  f(\bm z_{k+1}) < f(\bm z_{k}) + \delta.
\end{equation}
The tolerance $\delta$ is usually set as twice the approximate standard deviation of the noise in the objective function evaluation, which can be estimated by collecting several evaluations at the initial value of the parameters~\cite{Spall2000}.

Regardless of whether the step is accepted, the Hessian estimate $\mathcal{H}''_{k}$ from Eqs.~\eqref{eq:hessian_inertia} and~\eqref{eq:complex_hessian_inertia} must be updated at every iteration.

\subsubsection{Resampling}

This technique is also known as gradient or Hessian averaging. It consists in computing the random estimators for the gradient and Hessian $N_R$ times per iteration to perform the corresponding variable update using the average of these estimators. This practice is recommended in noisy environments~\cite{Spall2000}.

Note that the authors of QN-SPSA~\cite{QN-SPSA} implement resampling only for the Hessian estimator with the premise that evaluating the metric is cheaper than evaluating the objective function, which could lead to a better convergence rate with little increment on the experimental resources. However, here we stick to the convention stated by~\cite{Spall2000}, which is also implemented on Qiskit~\cite{Qiskit}.

\subsubsection{Post-processing}\label{postproces}

Several post-processing procedures have been proposed to improve the stability of the 2SPSA algorithm~\cite{Spall2007}. We consider two alternatives; the original proposal given by Eqs.~\eqref{eq:hessian}, and the procedure given by
\begin{subequations}
  \begin{align}
    \label{eq:hessian_hermitian_}
    \mathcal{H}_{k}' &= \frac{\mathcal{H}_{k} + \left[\mathcal{H}_{k}\right]^{\dagger}}{2},\\
    \label{eq:hessian_regularization_ours}
    \mathcal{H}_{k}'' &= \sqrt{\mathcal{H}_{k}'^{2} + \varepsilon I}, \\
    \label{eq:hessian_inertia_ours}
    \overline{\mathcal{H}}_{k} &= \frac{k}{k+1}\overline{\mathcal{H}}_{k-1} + \frac{1}{k+1}\mathcal{H}_{k}''.
  \end{align}
  \label{eq:Gidi}
\end{subequations}

\subsubsection{Scalar Preconditioning Approximation}
\label{sec:scalar-preconditioning}

Preconditioned methods, such as 2SPSA, 2CSPSA, QN-SPSA, and QN-CSPSA, adaptively adjust the descent direction and magnitude by adding a preconditioner to the stochastic approximation. However, these methods can exhibit numerical instabilities due to the inversion of a possibly ill-conditioned Hessian estimation. Postprocessing procedures can partially mitigate these issues, but these methods still lack consistency in numerical simulations compared to first-order methods. Most likely, these problems are caused by an inadequate adjustment of the descent direction. We consider these problems most likely induced by an inadequate adjustment of the descent direction.
 
It has been suggested~\cite{Zhu2002} to replace the Hessian estimation with a scalar function of its eigenvalues. Thereby, the descent direction is chosen according to the first-order gradient estimator while retaining the descent magnitude adaptivity from the preconditioner. Following these considerations, we propose a scalar approximation to the Hessian estimates \eqref{eq:hessian_approx_components}, \eqref{eq:QN_hessian_approx}, \eqref{eq:complex_hessian_approx}, and \eqref{eq:complex_QN_hessian_approx}. Specifically, we omit the stochastic perturbations $ \bm \Delta_k $ and $\bm{\tilde{\Delta}}_{k}$ presented in the Hessian estimates to only adjust the descent magnitude and preserve the first-order descent direction. Namely, we approximate the Hessian estimate of the second-order methods by
\begin{align}
  \label{eq:second_order_scalar_preconditioner}
  \mathcal{H}_{k}' = \frac{\delta^2f_k(\bm z)}{2b_k\tilde{b}_k},
\end{align}
and the Hessian estimates for quantum natural optimizers by
\begin{align}  \label{eq:quantum_natural_scalar_preconditioner}
  \mathcal{H'} &= -\frac{\delta^{2}F_{k}}{4b_{k}\tilde{b}_{k}}.
\end{align}

From this procedure, we consider a new set of second-order and quantum natural methods where the computational complexity is reduced. Namely, the number of classical operations on each iteration is reduced from $O(p^{3})$ to $O(p)$ where $p$ is the number of variables.

\section{Applications}\label{sec:appl}

We study the performance of the above optimization methods by comparing the rate of convergence of the objective function towards the minimum as a function of the number of iterations. We consider three important applications: variational quantum eigensolver, quantum control, and quantum state estimation. We use the variational quantum eigensolver to obtain the ground state energy of the Heisenberg Hamiltonian, which is a ubiquitous and relatively simple model that describes the interactions within a chain of spins. For quantum control, we implement the GRadient Ascent Pulse Engineering (GRAPE) method~\cite{QC-GRAPE}, which approximates a control pulse by a sequence of constant-intensity pulses. The control parameters of this pulse are optimized to find the best implementation of a given state, even in the presence of noise~\cite{PhysRevA.91.052306}. Finally, for quantum state estimation, we implement Self-Guided Quantum Tomography (SGQT)~\cite{PhysRevLett.113.190404}, based on minimizing the infidelity between an unknown state and a known parameterized state.

The studied optimization methods are stochastic. We use ensembles of numerical simulations to estimate the mean, standard deviation, median, and interquartile range of the objective function. Measurements are simulated by sampling a multinomial distribution with various numbers of trials. In the figures below, only the upper half of the standard deviation is shown.

We test the optimization methods considering different configurations and look for the ones that offer the best performance. The configurations we tested are all possible combinations of the following alternatives: with or without blocking, resampling with $N_{R}= 1, 2, 5$, the two basic post-processing procedures, Eqs.~\eqref{eq:complex_hessian} or Eqs.~\eqref{eq:Gidi}, and standard, asymptotic or static set of gain coefficients. The standard set is given by $a = 3$, $b = 0.1$, $A = 0$, $s = 0.602$ and $t = 0.101$, the asymptotic set by $a = 3$, $b = 0.1$, $A = 0$, $s = 1$ and $t = 1/6$, and the static set by $a = 0.01$, $b = 0.01$, $A = 0$, $s = 0$ and $t = 0$.

For clarity, we consider simulations with two groups of methods: (i) vanilla methods and (ii) improved methods, that is, the vanilla methods implemented with the improvements proposed in Subsec.~\ref{sec:improv}. The reason behind this separation lies in the drastic increase in resources required to perform blocking and resampling, and it could be useful to be able to discriminate when it is really worth swapping resources for better results.

We have created a freely available library~\cite{ComplexSPSA.jl} that contains the codes in the Julia programming language~\cite{Julia} that implements all of the optimization methods.

\begin{figure}[t]
  \centering
  \includegraphics[]{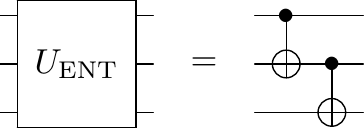}
  \caption{Entangling gate $U_\mathrm{ENT}$.
  \label{fig:ent-gate/vqe}}
\end{figure}

\begin{figure}[t]
  \centering    
  \includegraphics[]{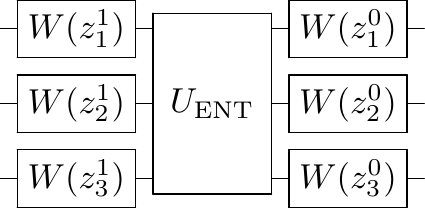}
  \caption{Parametric circuit used to implement VQE for
    the Heisenberg Hamiltonian of three qubits. $U_{\mathrm{ENT}}$ is
    an entangling gate depicted in Fig.~\ref{fig:ent-gate/vqe}.
    \label{fig:par-circuit/vqe}}
\end{figure}

\subsection{Variational Quantum Eigensolver}

The search for the ground state and its energy $E_0$ of a Hamiltonian is a problem of great interest in areas such as computational chemistry and condensed matter physics. This is because much of the phenomenology and properties of quantum systems can be studied from the ground state and its energy. However, finding this eigenstate in large systems is not a trivial task. It is often infeasible due to the exponential growth of the dimension of the Hilbert space with respect to the number of subsystems. For large systems, the Rayleigh-Ritz method~\cite{Ritz1909,Raylaigh1870} is a useful tool since it is limited to searching a parameterized subset of the original Hilbert space to reduce the computational cost of optimization. A further reduction in computational cost is achieved using the variational quantum eigensolver (VQE) method~\cite{Peruzzo2014}. This consists of performing the Rayleigh-Ritz method with the help of a classical and a quantum computer, which makes it a promising tool for the current generation of quantum technologies.

The goal is to find the eigenstate $|\psi_0\rangle$ associated with the lowest eigenvalue $E_0$ of a Hamiltonian. This ground state can be characterized as the solution to the optimization problem
\begin{equation}
  E_0=\min_{|\psi\rangle}
  \frac{\langle\psi|H|\psi\rangle}{\langle\psi|\psi\rangle}.
\end{equation}
The Rayleigh-Ritz method provides an estimate of $E_0$ by parameterizing the trial states as $|\psi(\bm{\theta})\rangle$ and optimizing over the vector $\bm{\theta}$ of parameters. The underlying idea is that the subset defined by the parameterization must have a smaller dimension than the total Hilbert space to reduce the computational cost.

The VQE method considers the generic hermitian Hamiltonian operator $H=\sum_{i=1}^nh_i\sigma_i$ and the trial states parameterization
\begin{equation}
  |\psi(\bm{\theta})\rangle=R_N(\theta_N)\cdots R_1(\theta_1)|\bm{0}\rangle,
\end{equation}
where $R_i(\theta_i)$ are quantum gates parameterized by $\theta_i$ and applied one after the other to the initial state $|\mathbf{0}\rangle$. This parameterization corresponds to a variational quantum circuit. The average energy $\langle\psi(\bm{\theta})|H|\psi(\bm{\theta})\rangle$
can then be computed by individually measuring each term $\langle\psi(\bm{\theta})|\sigma_i|\psi(\bm{\theta})\rangle$ on a quantum computer and adding the results weighted with their respective coefficients $h_i$. Thereafter, the values of $\langle\psi(\bm{\theta})|H|\psi(\bm{\theta})\rangle$ are used by a suitable optimization method running on a classic computer.

In general, the VQE method uses SPSA as the optimization algorithm due to its robustness against noise, which suggests that the optimization methods presented here can also be used. In order to evaluate the performance of these various methods, we use as a testing ground the problem of finding the ground state energy of the Heisenberg Hamiltonian, which models the magnetic interaction of a ferromagnetic lattice. This is given by the expression
\begin{equation}
  H_H = j\sum_{\langle m,n\rangle}\sum_{k=x,y,z} \sigma_m^k\sigma_n^k
  + h\sum_m \sigma_m^z,
  \label{eq:heisenberg/vqe}
\end{equation}
where $j$ and $h$ are dimensionless coupling constants between neighboring sites and with an external magnetic field, respectively, $\sigma_m^x,\sigma_m^y,\sigma_m^z$ are the single-qubit Pauli operators acting on the $m$-th lattice site, and $\langle m,n\rangle$ indicates that the sum is performed on the nearest neighbors in the lattice. To parameterize the trial states we use the complex parametric single-qubit gate
\begin{equation}
  W(z) = e^{-i( z\sigma_+ +z^{*}\sigma_-) },
\end{equation}
where $z$ is a complex parameter and $\sigma_\pm=\sigma^x\pm i \sigma^y$. This gate can be implemented experimentally by a sequence of three real parameter gates.

The parameterization used for the trial states is given by
\begin{align}
  |\psi(\mathbf{z})\rangle
  = \prod_{q=1}^N&W_{q}^{d}(z_q^d)U_{\mathrm{ENT}}\times\cdots\nonumber\\
                 &\times\prod_{q=1}^NW_{q}^{1}(z_q^1)U_{\mathrm{ENT}}
                   \prod_{q=1}^NW_{q}^{0}(z_q^0)|\mathbf{0}\rangle,
                   \label{eq:state_circuit/vqe}
\end{align}
where $W^l_q$ corresponds to applying $W$ on the $q$-th qubit, $l$ indicates the layer of the circuit, and $U_{\mathrm{ENT}}$ is the three-qubit entangling gate depicted in Fig.~\ref{fig:ent-gate/vqe}. Similarly for $z^l_q$.

To evaluate the performance of the different algorithms we consider the Heisenberg Hamiltonian Eq.~\eqref{eq:heisenberg/vqe} with $h=0.3$ and $j=1$ for a ring of $10$ qubits with periodic boundary conditions $q_{i+10} = q_i$. The trial states are parametrized by Eq.~\eqref{eq:state_circuit/vqe} with $d=1$ entangling layers, as depicted by the circuit in Fig.~\Ref{fig:par-circuit/vqe}. Each algorithm is simulated considering $10^2$ randomly selected initial states according to a Haar distribution, which allows estimating statistical indicators such as mean, median, standard deviation, and interquartile range. The measurements required by each optimization method are simulated with an ensemble size of $2\times10^4$. The standard gain coefficients used in first-order algorithms are $b=0.1$, and $a$ follows a calibration based on~\cite{Kandala2017}.

\begin{figure*}
    \centering
    \includegraphics[width=\textwidth]{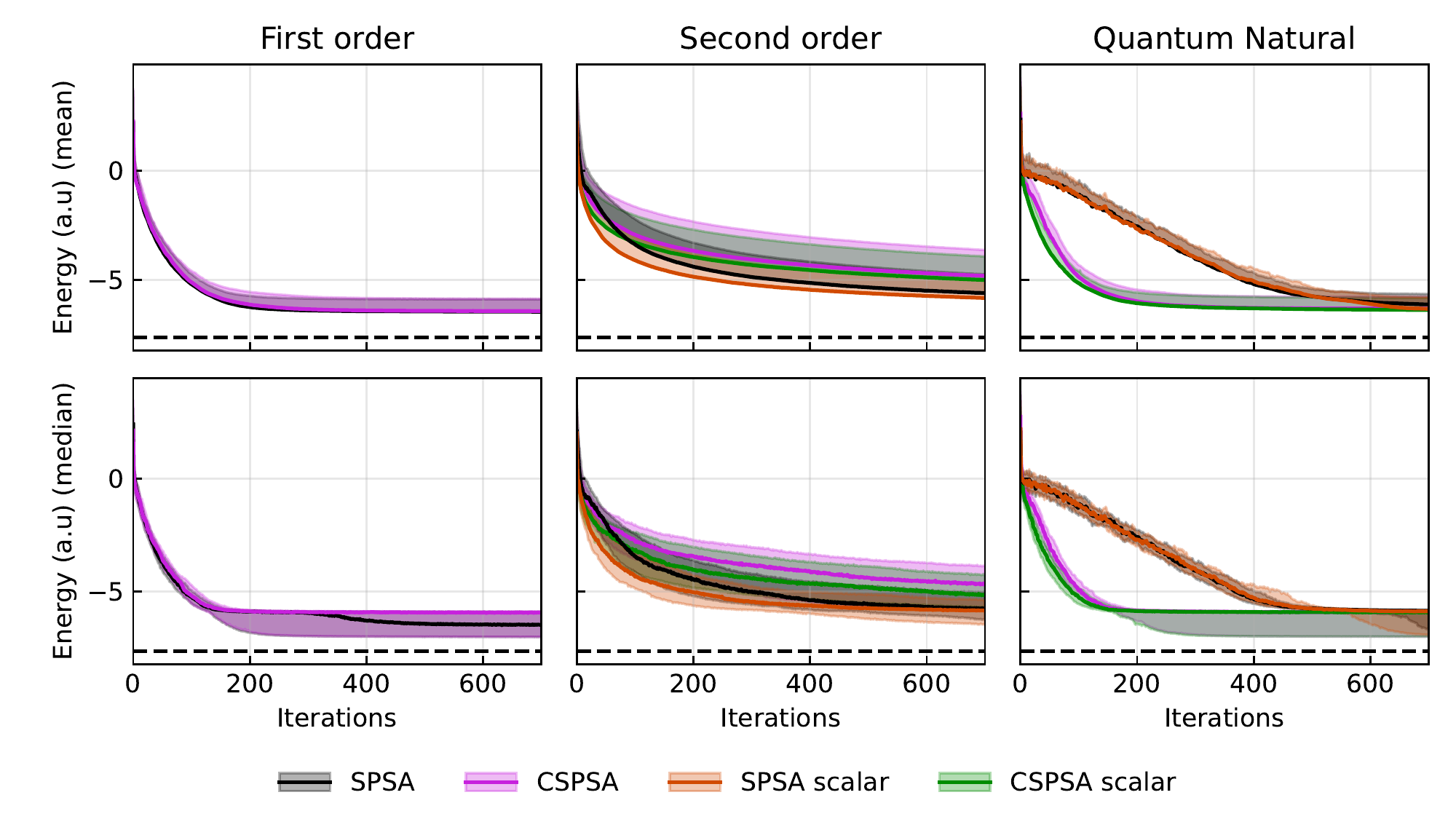}
    \caption{The mean (top row) and median (bottom row) of the energy (in arbitrary units) as a function of the number of iterations obtained through the VQE for the Heisenberg Hamiltonian in a 10-qubit ring configuration using vanilla optimization algorithms. The shaded areas represent the variance (top row) and the interquartile range (bottom row). The dashed line indicates the exact minimum. The statistics are obtained from a sample of $10^2$ randomly generated states to estimate the minimum energy. The measurements in each circuit were estimated with $2\times10^4$ shots. The values of the gain coefficients and post-processing class can be found in Table \ref{tab:vqe_vanilla} of Appendix \ref{sec: Tables}.}
    \label{fig:vanilla-plots/vqe}
\end{figure*}

Figure~\ref{fig:vanilla-plots/vqe} displays the best performance of the vanilla algorithms as a function of the number of iterations. All methods delivered the best results using the post-processing Eq.~\eqref{eq:Gidi}. This figure shows that the best performers are the real and complex first-order and the quantum-natural complex algorithms, which exhibit an almost indistinguishable behavior in mean, median, standard deviation, and interquartile range. These algorithms converge to a minimum at approximately $2\times10^2$ iterations, after which become approximately constant. The second-order algorithms exhibit a slower convergence, reaching a similar value only after $7\times10^2$ iterations.

\begin{figure*}
    \centering
    \includegraphics[width=\textwidth]{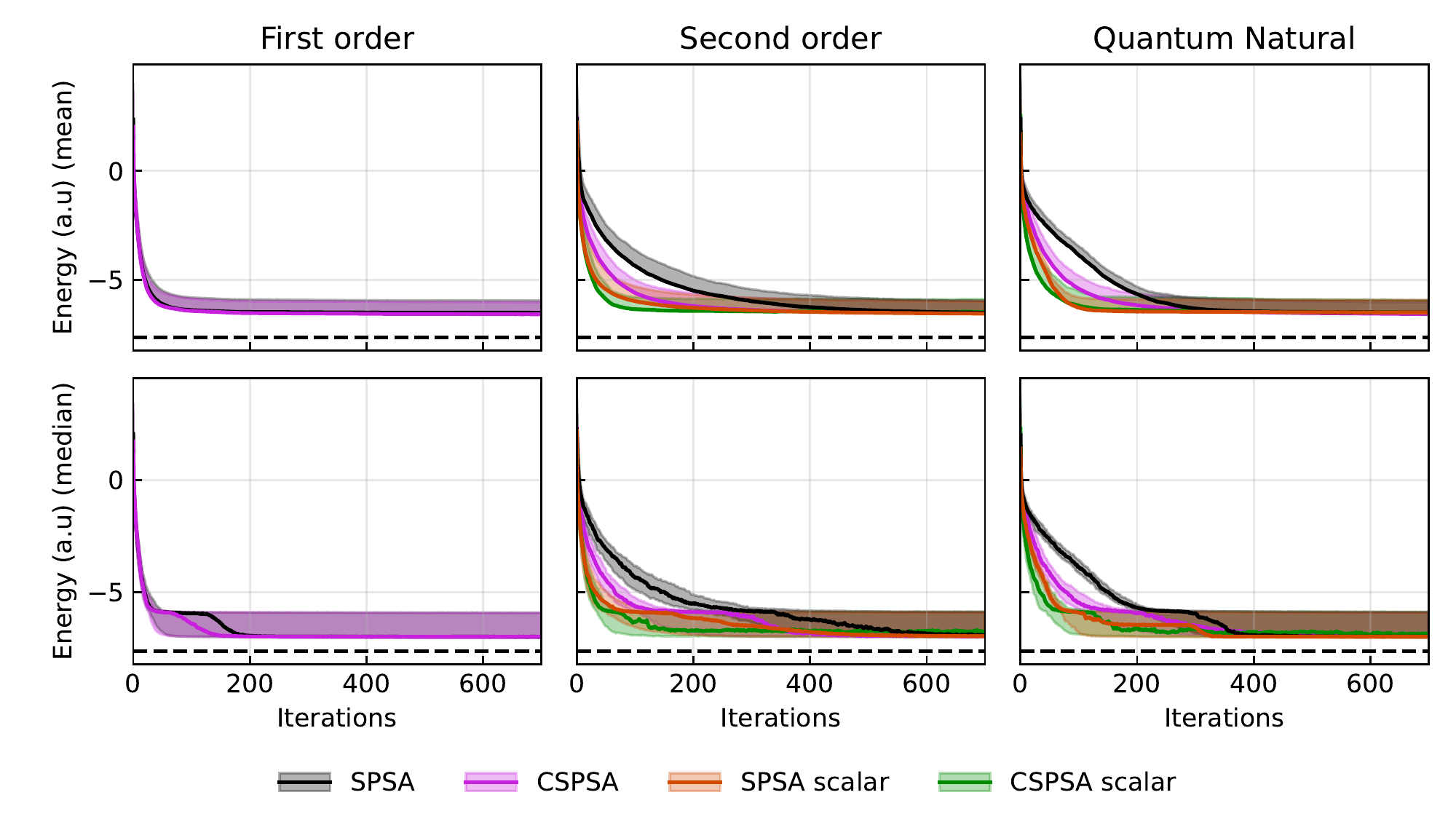}
    \caption{The mean (top row) and median (bottom row) of the energy as a function of the number of iterations obtained through the VQE for the Heisenberg Hamiltonian in a 10-qubit ring configuration using improved optimization algorithms. The shaded areas represent the variance (top row) and the interquartile range (bottom row). The dashed line indicates the exact minimum. The statistics are obtained from a sample of $10^2$ randomly generated states to estimate the minimum energy. The measurements in each circuit were estimated with $2\times10^4$ shots. The values of the gain coefficients, post-processing class, and the setting of resampling and blocking can be found in Table \ref{tab:vqe_best} of Appendix \ref{sec: Tables}.}
    \label{fig:improved-plots/vqe}
\end{figure*}

Figure~\ref{fig:improved-plots/vqe} displays the best performance of the improved methods. In this case, the best performers are the first-order methods, the scalar version of second-order methods, and the scalar version of quantum-natural methods. These exhibit an almost indistinguishable convergence in mean and median as well as similar dispersion. In particular, a minimum is achieved at approximately $\times10^2$ iterations after which the energy becomes nearly constant. 2CSPSA and 2SPSA scalar methods use post-processing of Eqs.~\eqref{eq:complex_hessian} and \eqref{eq:Gidi}, respectively, and standard gains. QN-CSPSA and QN-SPSA scalar methods use post-processing of Eqs.~\eqref{eq:complex_hessian} and \eqref{eq:Gidi}, respectively, and asymptotic gains. All best performers use resampling with $N_R=5$.

From Figs.~\ref{fig:vanilla-plots/vqe} and \ref{fig:improved-plots/vqe} we conclude that the best performance in the variational quantum eigensolver is achieved using SPSA, CSPSA, QN-CSPSA, and QN-CSPSA scalar methods, which does not significantly differ in their vanilla or improved versions. Blocking and resampling lead to a clear improvement of the second-order methods, delivering results similar to the best performers.

First-order algorithms provide the best performance for this particular problem. Nonetheless, it's crucial to note that this level of performance was attained through a resource-intensive search for gain coefficients. In the absence of such a search, the first-order algorithms performed poorly. To bypass the calibration of the gain coefficients, quantum natural algorithms can be applied while achieving a performance close to the calibrated CSPSA. In particular, the scalar quantum natural CSPSA algorithm also reduces the classical computational cost.

\subsection{Quantum Control} 
Quantum control theory lays a firm theoretical foundation for developing a series of systematic methods that allow the manipulation and control of quantum systems. In particular, the search for an optimized time evolution that allows guiding the system from an initial state to a desired final state is of great interest. Quantum control theory has already achieved significant successes in physical chemistry~\cite{QC-Quemistry1}, atomic and molecular physics~\cite{QC-AtomicMolecular}, quantum optics~\cite{QC-QuantumOptics}, and has also contributed to understanding fundamental aspects of quantum mechanics~\cite{QC-QM}. In recent years, the development of the general principles of quantum control theory has been recognized as an essential requirement for the current and future applications of quantum technologies.

A particular problem in quantum control is the precise engineering of quantum states, that is, whether a quantum system can be driven into a given state. This problem has practical importance since it is closely related to the universality of quantum computing and the possibility of achieving transformations at the atomic or molecular scale. An important research problem is that of finite-dimensional quantum systems, for which the controllability criteria can be expressed in terms of parameters included in the Hamiltonian of the system.

The quantum state control problem~\cite{QC-QOCT} consists in identifying an appropriate set of time-dependent control parameters $u_{k}(t)$ in such a way that its controlled change in time guides the evolution of the system from an initial state $\ket{\psi_{0}}$ to a predetermined objective state $\ket{\psi_{f}}$. The control parameters enter in the Hamiltonian as coefficients in a linear combination of operators, that is,
\begin{equation}
  H(t)= H_{0}+ \frac{1}{2}\sum_{k}\left(u_{k}(t)C_{k}+u^{*}_{k}(t)C^{\dagger}_{k}\right),
  \label{eq:QC2}
\end{equation}
where the set $\{C_k\}$ are a base of operators and we allow the possibility of complex control parameters. In order to obtain this set of parameters, it is necessary to solve the time-dependent Schr\"odinger equation. Unfortunately, solutions of the Schr\"odinger equation for a time-dependent Hamiltonian cannot generally be obtained analytically. However, it is possible in certain cases to use techniques developed in the area of adiabatic control ~\cite{Burgarth2019,PhysRevA.99.062306,Zhou2016,PhysRevA.72.012114}.

To overcome this problem, we use the GRadient Ascent Pulse Engineering (GRAPE) method~\cite{QC-GRAPE}, originally introduced in nuclear magnetic resonance spectroscopy and proposed to design a pulse sequence that drives the evolution toward the optimum of a predefined objective function. This method allows us to compute the evolution of a time-dependent Hamiltonian through a sequence $H_m$ of time-independent Hamiltonians. The total evolution time $T$ is divided into a number $M$ of time intervals $\Delta t_m=t_{m+1}-t_m$ ($m=0,\dots,M-1$), which are normally of equal length so that in each interval the control parameters $u_{k}(t)$ are approximately constant. In each time interval, the evolution is given by
\begin{equation}
  U_m = e^{-i\Delta t_mH_m},
  \label{eq:QC3}
\end{equation}
where $H_{m}=H(t^*_{m})$ with $t^*_m\in[t_m,t_{m+1}]$. A classical optimization algorithm is used to obtain the values of the control parameters that lead to the optimum of the objective function. The evolution of the system at time $T$ is thus approximated by the sequence
\begin{equation}
  U_{\text{GRAPE}}= \prod_{m=0}^{M-1}U_{m}
  \label{eq:QC4}
\end{equation}
and the state of the system at time $T$ is
\begin{equation}
  \ket*{\tilde{\psi}_{f}}= U_{\text{GRAPE}}\ket{\psi_{0}}.
  \label{eq:QC5}
\end{equation}
Once a propagator has been computed for a set of control parameters, all that remains is to choose an objective function to compare the target state with the state given by the evolution for a given set of control parameter values. In our case, we use the infidelity that is given by
\begin{equation} 
  I(|\tilde\psi_{f}\rangle,|\psi_{f}\rangle)= 1 - |\braket*{\tilde\psi_{f}}{\psi_{f}}|^{2},
  \label{eq:QC6}
\end{equation}
which is minimized with an optimization algorithm. The original GRAPE proposal uses the descending gradient algorithm. The dimension of the search space is given by $N_pM$, where $N_p$ is the number of parameters, and therefore can be very large.

To test the optimization methods introduced here, we turn to the quantum control of a five-qubit system, where we aim at preparing the target state $\ket{\psi_{f}}= \ket{0}^{\otimes 5}$ by controlling the evolution generated by the Heisenberg Hamiltonian given by
    \begin{equation}
        H_H(t)=-\frac{1}{2}\sum_{k=x,y,z}J_{k}(t)\sum_{\langle m,n\rangle}
        \sigma_{m}^{k}\sigma_{n}^{k},
        \label{eq:QC7}
    \end{equation}
which depends on the three complex coupling constants $J_{x}(t), J_{y}(t)$ and $J_{z}(t)$. These play the role of control parameters whose values are driven by the quantum control method to approach the desired target state.

After applying the GRAPE method for the evolution of the system, the final state is
    \begin{equation}
        \ket*{\tilde\psi_{f}}= \prod_{i=1}^{M-1} e^{-i\Delta t_mH_{H}(t^*_m)}\ket{\psi_{0}} ,
        \label{eq:QC8}
    \end{equation}
where $\ket{\psi_{0}}$ is an initial five-qubit state and $H_H$ contains the control parameters.

Our performance study is based on numerical simulations where we implement GRAPE with each of the methods reviewed or proposed here. For a given optimization method, we start by choosing an initial state $\ket{\psi_{0}}$ from a Haar-uniform distribution on which we apply the GRAPE method with $M=25$ and $10^3$ iterations. Therefore, the dimension of the complex search space is $75$, with the real search space being twice as large. The measurements required by the optimization method are simulated with an ensemble of size $2^{13}$. This procedure is repeated $10^4$ times to obtain estimates of relevant statistical indicators such as mean, median, standard deviation, and interquartile range, as functions of the number of iterations. The gain parameters used in the numerical simulations are shown in TABLES~\ref{tab:qc_vanilla} and~\ref{tab:qc_best} of Appendix~\ref{sec: Tables}.

\begin{figure*}
  \centering
  \includegraphics[width=\textwidth]{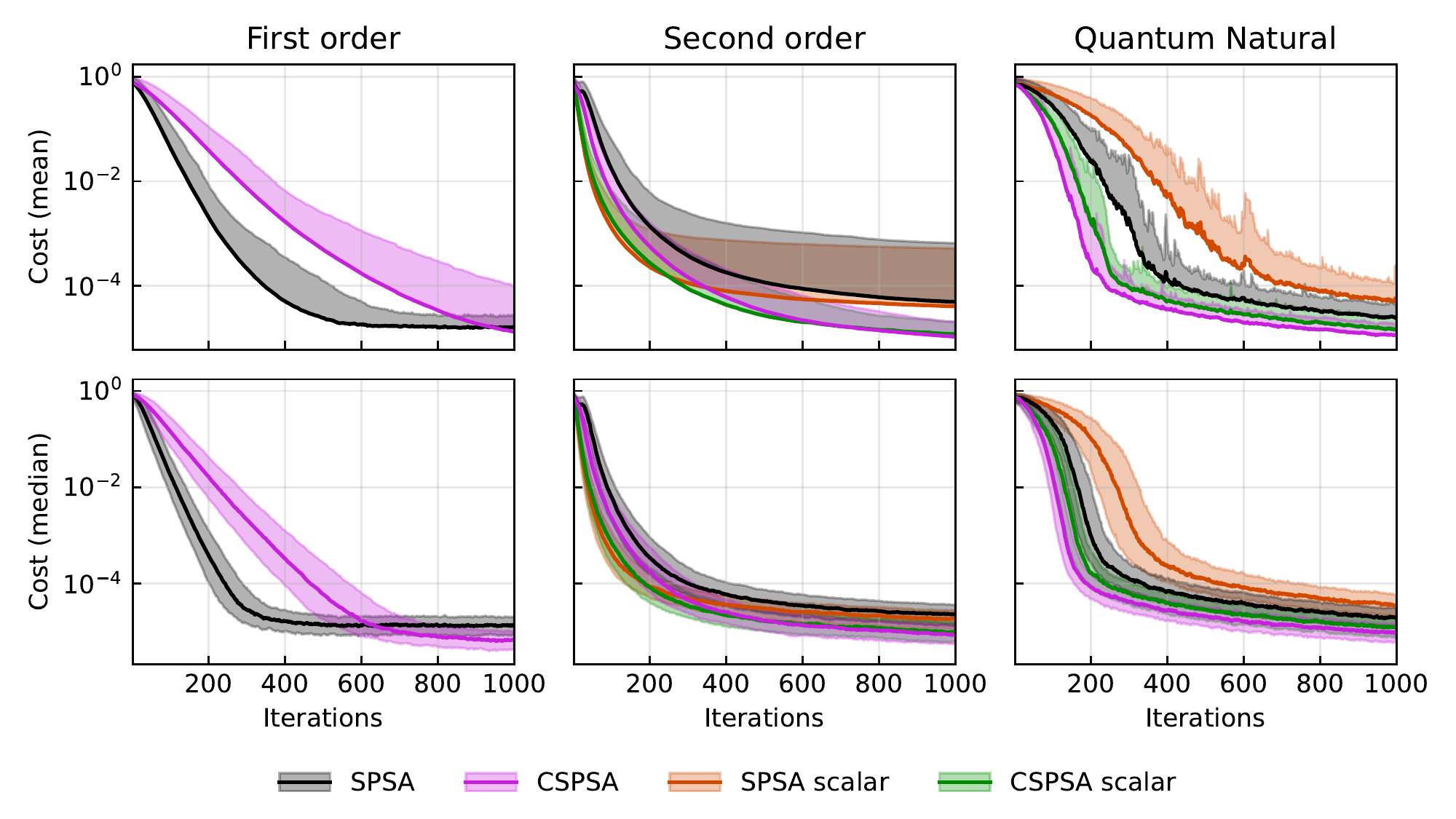}
  \caption{The mean (top row) and median (bottom row) of the infidelity as a function of the number of iterations obtained through the GRAPE method applied to the quantum control of a 5-qubit pure state and vanilla optimization algorithms. The shaded areas represent the variance (top row) and the interquartile range (bottom row). The values of the infidelity are obtained by simulating a measurement process with a sample size of $2^{13}$ and $25$ iterations of GRAPE, $10^{4}$ shots per measurement, and $10^3$ iterations, which are generated through uniformly distributed initial states $\ket{\psi_{0}}$. The values of the gain coefficients and post-processing class can be found in Table \ref{tab:qc_vanilla} of Appendix \ref{sec: Tables}.}
  \label{fig:VanilaQC}
\end{figure*}

\begin{figure*}
    \centering
    \includegraphics[width=\textwidth]{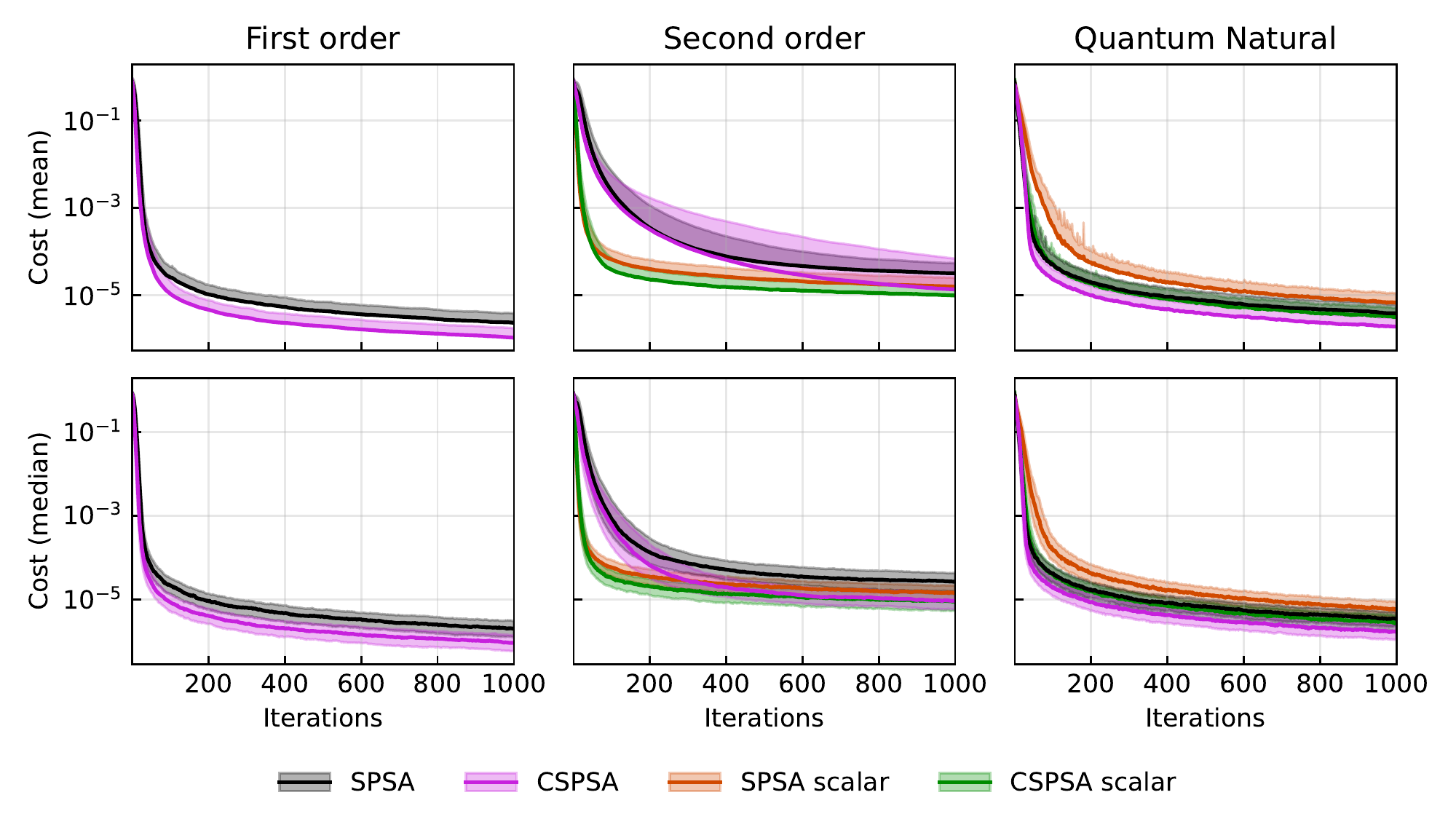}
    \caption{The mean (top row) and median (bottom row) of the infidelity as a function of the number of iterations obtained through the GRAPE method applied to the quantum control of a 5-qubit pure state and improved optimization algorithms. The shaded areas represent the variance (top row) and the interquartile range (bottom row). The values of the infidelity are obtained by simulating a measurement process with a sample size of $2^{13}$ and $25$ iterations of GRAPE, $10^{4}$ shots per measurement, and $10^3$ iterations, which are generated through uniformly distributed initial states $\ket{\psi_{0}}$. The values of the gain coefficients, post-processing class, and the setting of resampling and blocking can be found in Table \ref{tab:qc_best} of Appendix \ref{sec: Tables}.}
    \label{fig:BestQC}
\end{figure*}

The results of the numerical simulations of the GRAPE method with the different optimization algorithms in the five-qubit case are depicted in Figs.~\ref{fig:VanilaQC} and \ref{fig:BestQC}, which show the best results among the vanilla methods and the improved methods, respectively. Each figure shows the value of the mean (upper row) and median (lower row) infidelity as a function of the number of iterations together with the variance (upper row) and the interquartile range (lower row) as shaded areas.

Figure \ref{fig:VanilaQC} shows the comparison between methods without using blocking and resampling (see \ref{sec:improv}), that is, the vanilla methods. Second-order methods exhibit the best mean performance, particularly 2CSPSA and scalar 2CSPSA. These are closely followed by their quantum natural counterparts. First-order methods initially offer a better convergence rate but stagnate after a certain number of iterations. Let us note that this is the only case among all applications where first-order SPSA and CSPSA achieve their best performance using the static gain coefficients. In the median, second-order methods exhibit a higher convergence rate, closely followed by the quantum natural methods. While first-order methods require a larger number of iterations, a similar value of infidelity is reached in all cases. However, they behave differently in mean and median, in contrast to the complex second-order and quantum natural methods exhibiting consistent statistical indicators.

Figure \ref{fig:BestQC} shows the comparison between methods when we allow the usage of blocking and resampling, that is, the improved methods. The best performance in the mean and the median is attained by the first-order methods seconded by the quantum natural methods, which exhibit a slightly slower rate of convergence with similar standard deviation and interquartile range. Second-order methods are the worst performers. These are characterized by a lower precision in mean, large standard deviation, and a slower rate of convergence, with the exception of scalar methods. Figure \ref{fig:BestQC} also indicates that complex methods perform better than their real counterparts. 

Generally, the first-order CSPSA method with resampling and blocking obtains the best result, using Eqs.~\eqref{eq:Gidi} for post-processing, closely followed by the QN-CSPSA method.

\subsection{Quantum state estimation}

Born's rule endows quantum mechanics with predictive power. According to this rule, the probability $p_k$ of obtaining a result $k$ in an experiment described by a POVM $\{E_k\}$ when the quantum system is described by a quantum state $\rho$ is given by the Hilbert-Schmidt inner product $p_k= \operatorname{Tr}(\rho E_k)$. Therefore, the comparison between the theoretical predictions and the experimental results requires an accurate characterization of the quantum state $\rho$ and of the experiment through the POVM $\{E_k\}$. This leads to the problem of estimating quantum states and processes. To do this, several quantum state estimation methods have been designed, most of them based on the post-processing of experimental data acquired through the measurement of a fixed informationally complete POVM. Adaptive measurements have also been used to design quantum state estimation methods. Today, methods for estimating quantum states are an important tool for both quantum communications and quantum computing and have been used for the characterization of single-photon and continuous variable states~\cite{Laiho2012,Braczyk2012,Mller2012,Chiuri2012,Wallentowitz2012}, cavity fields~\cite{Sayrin2012}, atomic ensembles~\cite{Christensen2013,ReydeCastro2013,Mitchell2012}, trapped ions~\cite{Gu2012,Hffner2005}, optical detectors~\cite{Zhang2012,Brida2012,Anis2012}, and for quantum key distribution~\cite{Watanabe2008}.

Recently, the estimation of finite-dimensional pure unknown states has been formulated as an optimization problem~\cite{PhysRevLett.113.190404}. According to this, the unknown state is characterized as the minimizer of infidelity $I\qty(\ket{\psi}, \ket{\phi}) = 1-\qty|\braket{\psi}{\phi}|^2$, that is,
\begin{equation}
  |\psi\rangle= \arg \qty(\min_{|\phi\rangle\in\cal{H}}I\qty(\ket{\psi}, \ket{\phi})).
\end{equation}
This suggests using optimization algorithms to minimize fidelity and estimate the unknown state $|\psi\rangle$, which has been called self-guided quantum tomography (SGQT). Gradient-based optimization is ruled out since it is not known how to measure the infidelity gradient with respect to the parameters entering the $|\phi\rangle$ state. However, infidelity can be measured by projecting the unknown state onto any basis containing the state $|\phi\rangle$. In this scenario, the optimization methods presented in the previous section can be used to experimentally implement the infidelity minimization according to SGQT. Initially, SGQT was based on the SPSA algorithm. Subsequently, CSPSA was introduced in SGQT, obtaining an improvement in the rate of convergence and a lower dispersion in the sample of estimates. More recently, CSPSA was combined with maximum likelihood estimation to achieve precision close to the lower limit of Gill-Massar, which is the best achievable estimation accuracy for pure states. Estimating pure states through SPSA and CSPSA has already been experimentally demonstrated~\cite{PhysRevLett.117.040402}.

We use pure state estimation through SGQT to test the performance of the optimization methods proposed in the previous sections. After selecting a particular optimization method, we generate an unknown 6-qubit pure state and an initial guess state from a Haar-uniform distribution. The optimization method is iterated $5\times10^3$ times and the fidelity values are obtained by simulating a measurement with binomial distribution on an ensemble size $N=2\times10^4$. This procedure is repeated $10^2$ times to generate estimates of relevant statistic indicators. The gain parameters used in the numerical simulations of each method are shown in Appendix \ref{sec: Tables}.

\begin{figure*}[ht!]
    \includegraphics[width=1.\textwidth]{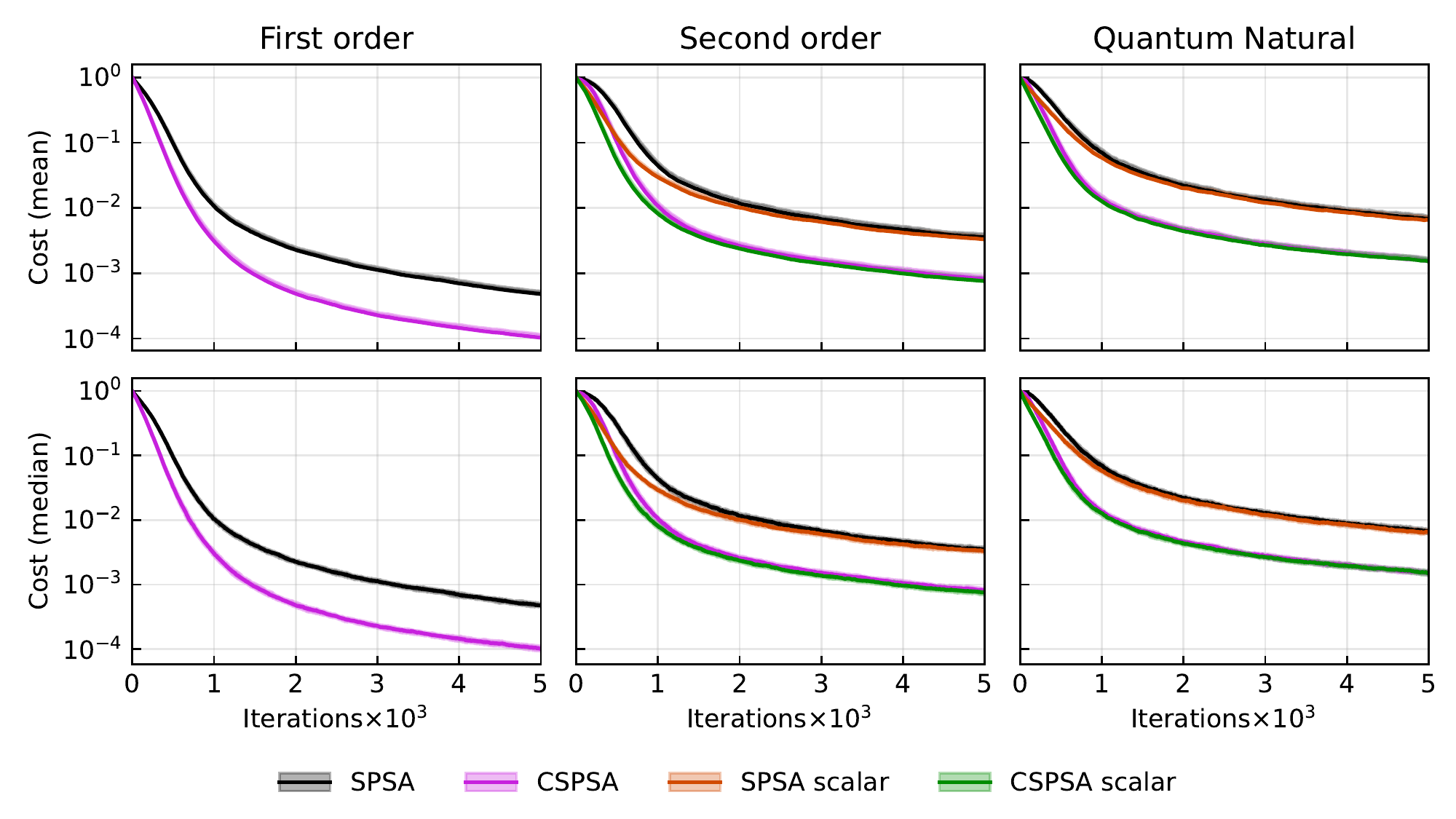}
\caption{The mean (top row) and median (bottom row) of the infidelity as a function of the number of iterations obtained by using SGQT to estimate six-qubit states and vanilla optimization algorithms. Shaded areas represent variance (top row) and interquartile range (bottom row). Statistical indicators are calculated from a sample of $10^2$ Haar-uniform distributed pairs of unknown and initial guess states. Measurements of the infidelity are simulated with a binomial distribution with $N=2\times10^4$ shots. The values of the gain coefficients and post-processing class can be found in Table \ref{tab:sgqt_vanilla} of Appendix \ref{sec: Tables}.}
\label{fig:SGQT_vanilla}
\end{figure*}

\begin{figure*}[ht!]
    \includegraphics[width=1.\textwidth]{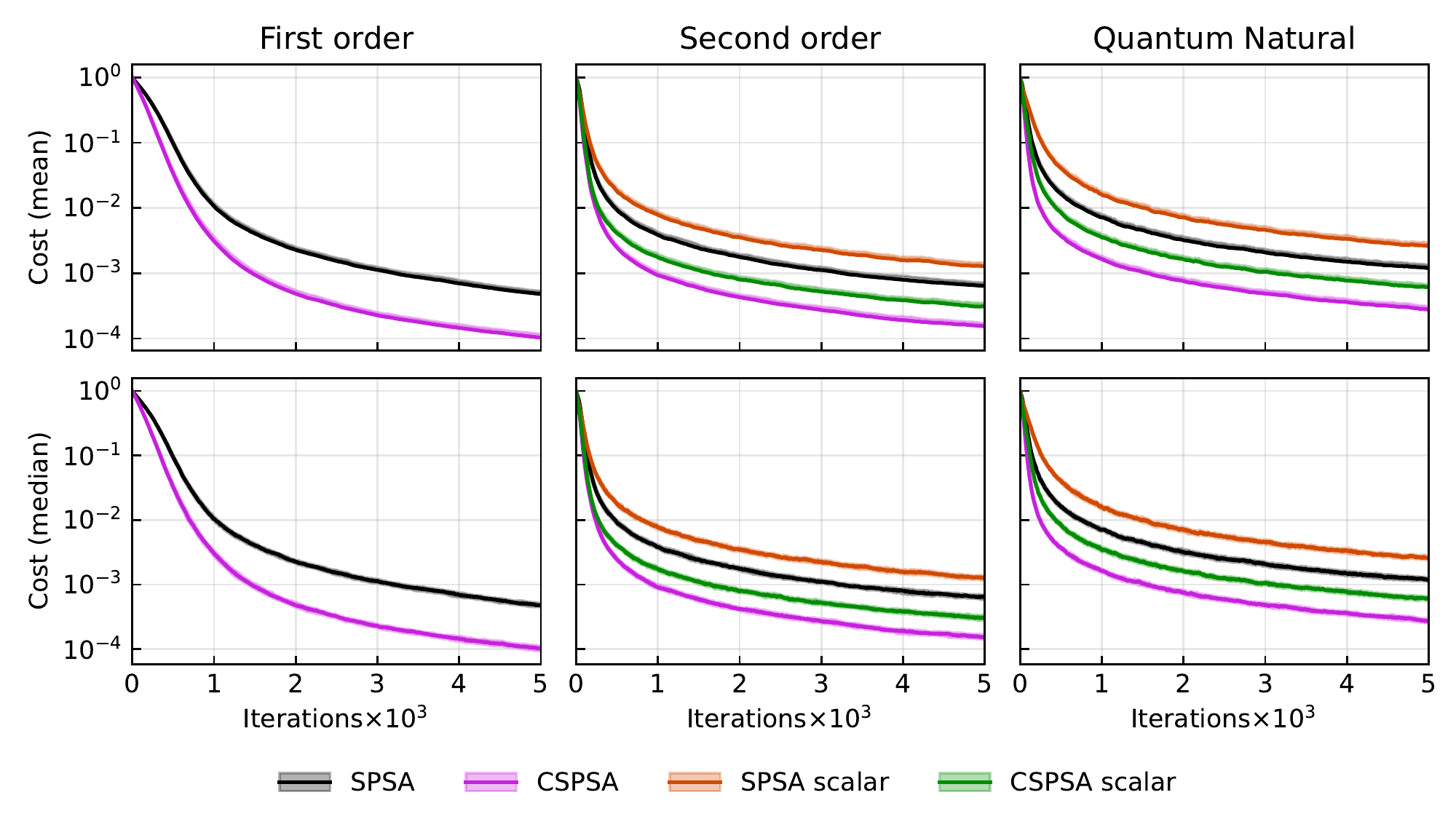}
\caption{The mean (top row) and median (bottom row) of the infidelity as a function of the number of iterations obtained by using SGQT to estimate six-qubit states and improved optimization algorithms. Shaded areas represent variance (top row) and interquartile range (bottom row). Statistical indicators are calculated from a sample of $10^2$ Haar-uniform distributed pairs of unknown and initial guess states. Measurements of the infidelity are simulated with a binomial distribution with $N=2\times10^4$ shots. The values of the gain coefficients, post-processing class, and the setting of resampling and blocking can be found in Table \ref{tab:sgqt_best} of Appendix \ref{sec: Tables}.}
\label{fig:SGQT_best}
\end{figure*}

The results of the numerical simulations of SGQT with the different optimization algorithms are depicted in Figs.~\ref{fig:SGQT_vanilla} and \ref{fig:SGQT_best} that show the best results among the vanilla and improved methods, respectively. Each figure shows the value of the mean (upper row) and median (lower row) infidelity as a function of the number of iterations together with the variance (upper row) and the interquartile range (lower row) as shaded areas. In every figure, the first column contains the results of SPSA and CSPSA. The second column contains the results obtained by the second-order algorithms, that is, 2SPSA, 2CSPSA, scalar 2SPSA, and scalar 2CSPSA. The third column contains the results obtained by the algorithms based on the quantum natural method, that is, QN-SPSA, QN-CSPSA, scalar QN-SPSA, and scalar QN-CSPSA.

In Figs.~\ref{fig:SGQT_vanilla} and \ref{fig:SGQT_best} mean and median values for each algorithm are very close. Furthermore, the variance and interquartile range are very narrow, which shows the absence of outliers in the generated samples. Typically, all optimization algorithms are characterized by a sharp decrease in infidelity followed by an approximately linear asymptotic regime.

Figure \ref{fig:SGQT_vanilla} shows the comparison between every method without using the blocking and resampling improvements in Eqs.~\eqref{eq:Gidi}, as these methods largely increase the number of resources. The first-order methods offer the best performance, getting about an order of magnitude improvement over the other methods. In contrast, the second-order methods perform slightly better than their QN counterpart. The scalar approximation shows no improvements for the second-order and quantum natural methods. The complex methods show better convergence than their real counterparts by about an order of magnitude.


Figure \ref{fig:SGQT_best} shows the comparison between methods when we allow the use of blocking and resampling. For first-order algorithms, gradient blocking and resampling show no improvement, while second-order and QN methods improve when the Hessian approximation is averaged $5$ times per iteration. This improvement decreases the performance difference between the first-order and the other methods. Second-order methods still perform slightly better than QN methods. We also note that the improvement obtained by resampling is smaller for the scalar approximation.

In our simulations, blocking does not improve our results when considering the Hessian post-processing Eqs.~\eqref{eq:Gidi}. On the other hand, when considering the post-processing Eqs.~\eqref{eq:hessian}, the blocking show great improvement which matches our median results, but with worse mean performance and with wider data variability (See Appendix \ref{sec: Tables}).

From Figs.~\ref{fig:SGQT_vanilla} and \ref{fig:SGQT_best} we conclude that the first-order methods show better mean and median performance in the estimation of pure 6-qubit states via SGQT, even without considering gradient resampling. In this scenario, second-order methods are not expected to work properly since the fidelity Hessian vanishes for pure states. This issue could lower both precision and convergence speed. However, the Hessian post-processing allows us to mitigate this problem by perturbing the Hessian matrix with a weighted identity matrix. In this way, the best result achieved by the second-order methods uses the post-processing Eqs.~\eqref{eq:Gidi}. Quantum natural-based methods show similar behavior, albeit with slightly slower convergence.

First-order methods perform the best even without considering gradient resampling. In contrast, second-order and quantum natural methods need resampling improvement to stay competitive but require a much higher number of resources.

\section{Conclusions}

In this article, we have exhaustively compared different stochastic optimization methods applied to real-valued functions of complex variables. We started by reviewing the theory of the SPSA algorithm and two of its variants: 2SPSA and QN-SPSA. These three methods use a simultaneous perturbation stochastic approximation of the gradient of the objective function to optimize it. SPSA is a first-order algorithm, while 2SPSA is a second-order algorithm. QN-SPSA is a first-order algorithm that preconditions considering a metric natural for the problem at hand. We also reviewed the CSPSA algorithm, which optimizes real functions of complex variables without resorting to the real and imaginary parts of complex variables. This is a more natural approach in quantum mechanics, where most functions have complex arguments. Using CSPSA as starting point, we proposed two new optimization methods: 2CSPSA and QN-CSPSA, which are the complex field formulations of their real counterparts.

All the optimization methods presented here share the property that the number of evaluations (or measurements) of the objective function does not depend on the dimension of the optimization problem. This is an important advantage when the number of parameters on which the objective function depends is large. The number of objective function evaluations is constant at each iteration but different for each method. SPSA and CSPSA use $2$ evaluations of the objective function per iteration. 2SPSA and 2CSPSA use $4$ evaluations of the objective function since they are second-order methods. Finally, QN-SPSA and QN-CSPSA use $2$ evaluations of the objective function plus the calculation of an approximation of a metric. If the metric is the Fubiny-Study metric tensor, then the approximation is calculated by evaluating the fidelity with respect to 4 different pure states.

To assess the performance of the optimization methods, we have compared them in three important applications in quantum computing:  variational quantum eigensolver applied to the Heisenberg Hamiltonian of a 10-qubit ring, quantum control applied to a 5-qubit pure quantum state, and quantum state estimation to reconstruct a 6-qubit pure quantum state. These three applications have different objective functions that need to be measured in a quantum device and iteratively optimized to obtain a solution. In particular, we have compared the convergence rate as a function of the number of iterations. To do this, we have considered vanilla and improved algorithms versions.

Our simulations show several interesting results. The best performance is systematically achieved by the first-order CSPSA algorithm. In the case of the variational quantum eigensolver, improved first-order CSPSA and SPSA algorithms provide the best performance, exhibiting identical mean and median and similar standard deviation and interquartile range. 
In quantum control, improved CSPSA achieves better convergence in mean and median than all other algorithms, exhibiting a narrow standard deviation and interquartile range. This is also the case for state estimation, although in this case, the vanilla version of the CSPSA algorithm is almost indistinguishable from its improved version. 

The second-best overall performance has mixed results. In the variational quantum eigensolver, the improved second-order and improved quantum natural algorithms lead to an almost indistinguishable performance, while in quantum control the improved quantum natural algorithms, particularly improved QN-CSPSA, are clearly second best. In this case, scalar second-order algorithms perform, in mean and median, similarly to quantum natural algorithms. In contrast, their non-scalar counterparts show much lower mean performance, indicating the presence of a large number of outliers. In the case of quantum state estimation, improved second-order algorithms provide better performance than their quantum natural counterparts. In particular, second-order CSPSA achieves the second-best performance.

Generally, vanilla second-order algorithms lead to lower performance than vanilla quantum natural algorithms. This is mitigated by blocking and resampling in the improved versions of second-order algorithms, which offer performance close to that of improved quantum natural algorithms. Furthermore, complex algorithms perform better than their real counterparts, although the difference may be statistically insignificant in certain cases.

While optimizing a function, it may be possible that no information about the Hessian matrix is available a priori, either because of its high complexity or because it cannot be easily obtained analytically or numerically. For such cases, it would be desirable that second-order methods, which are based on the Hessian matrix approximation, would still be useful in the event that the Hessian matrix exhibits singularities. This is the case of quantum state estimation, where the Hessian vanishes identically. Nevertheless, second-order methods display a performance similar to first-order methods. Hessian post-processing Eqs.~\eqref{eq:Gidi} ensures that the preconditioning matrix is proportional to the identity for a vanishing Hessian matrix. This leads to second-order methods working like first-order methods, albeit possibly with sub-optimal gain coefficients.

The stochastic optimization methods studied here are defined through a set of gain parameters whose values specify the gain coefficients. These in turn control the step size and magnitude of the approximation of the gradient. In this way, the gain parameters are hyper-parameters that allow controlling the algorithms' convergence rate. In principle, it is conceivable to find gain parameters that lead to the best convergence rate. This is, however, an expensive optimization problem whose solution might even depend on the optimizer of the objective function. Therefore, it is usual to resort to gain parameters that have proven to be good enough in practice. We have resorted to the standard gain parameters, which lead to a fast convergence in the regime of a small number of iterations, and to the asymptotic gain parameters, which lead to a fast convergence in the regime of a large number of iterations. Let us note that a change in the gain parameters affects not only the mean and median convergence but also the variance and interquartile range. We have also performed our simulations considering static gain coefficients, which only led to a significant improvement in the case of vanilla first-order methods applied to quantum control.

From numerical simulations with fewer qubits, we observed that the performance difference between quantum natural and first-order algorithms tends to narrow as the number of qubits increases. For the simulations reported here, the performance difference among these algorithms is small. This may indicate that quantum natural methods may outperform first-order methods for a larger number of qubits. However, this advantage of quantum natural methods is obtained by increasing the number of measurements and the classical computational cost. In this scene, the scalar quantum natural methods proposed here might be a good alternative since, according to our results, they offer comparable performance at a reduced classical cost.

According to the applications considered here, vanilla first-order algorithms are efficient and reliable options for the most general case. If higher accuracy is needed, improved first-order algorithms are the straight choice. First-order methods may require careful calibration of the gain parameters, in which case the quantum natural algorithms are a suitable alternative. In addition, quantum natural algorithms show promising results for many qubits, while second-order algorithms do not exhibit a comparative advantage.

In our study of first- and second-order algorithms, we have considered a single source of noise, namely, the statistical character of quantum measurements. It is possible to consider other error sources, such as those affecting NISQ processors. However, first-order algorithms, real or complex, have convergence proofs that allow for certain types of errors affecting the evaluation of the target function. Thereby, it is expected that these algorithms will converge even in the presence of moderate noise, albeit with an increased number of iterations. The scenario in the case of the preconditioned algorithms is less clear due to the inversion of the approximated Hessian matrix. Therefore, a natural extension of this work would be to consider realistic noise sources and their impact on the convergence rate. Also, we have considered the performance as a function of the number of iterations. It is possible, however, to consider other valuable resources such as the number of measurements, evaluations, and circuits. These should also be considered in further studies of the real performance of optimization algorithms.

\begin{acknowledgments}
This work was supported by ANID -- Millennium Science Initiative Program -- ICN17$_-$012. JG was supported by ANID Chile, National Doctoral Degree Scholarship No. 21202616. LP was supported by ANID-PFCHA/DOCTORADO-BECAS-CHILE/2019-772200275, the CSIC Interdisciplinary Thematic Platform (PTI+) on Quantum Technologies (PTI-QTEP+), and the Proyecto Sinérgico CAM 2020 Y2020/TCS-6545 (NanoQuCo-CM). LZ was supported by ANID-PFCHA/DOCTORADO-NACIONAL/2018-21181021, the Government of Spain (Severo Ochoa CEX2019-000910-S, TRANQI and European Union NextGenerationEU PRTR-C17.I1), Fundació Cellex, Fundació Mir-Puig and Generalitat de Catalunya (CERCA program). AD was supported by ANID Grants 1231940 and 1230586.

\end{acknowledgments}

\bibliography{mainbio}

\newpage
\onecolumngrid
\appendix

\section{Tables}\label{sec: Tables}

In this section, we provide the value of the statistical indicators; median, inter-quartile range (IQR), mean, and standard deviation (STD), obtained through numerical simulations for the best configuration of each optimization method on each application. These values were used to determine the algorithm with the best performance after $700$, $1000$, and 
$5000$ iterations for the variational quantum eigensolver, quantum control of quantum states, and self-guided quantum tomography, respectively. We indicate the gain coefficients and equations used for post-processing for each vanilla method. In the case of improved methods, we also indicate the amount of resampling and the use of blocking.

\begin{table}[ht!]
  \begin{tabular}{|c|c|c|c|c|c|c|}
    \hline
    {\bf Method} & {\bf Gains} & {\bf Post-processing} & {\bf Median} & {\bf IQR} & {\bf Mean} & {\bf STD} \\
    \hline
    SPSA & Standard & - & $-6.48$ & $1.11$ & $-6.46$ & $0.56$ \\
    \hline
    CSPSA & Standard & - & $-5.93$ & $1.12$ & $-6.44$ & $0.57$ \\
    \hline
    2SPSA & Standard & Eqs.~\eqref{eq:Gidi} & $-5.76$ & $1.17$ & $-5.61$ & $0.85$ \\
    \hline
    2CSPSA & Standard & Eqs.~\eqref{eq:Gidi} & $-4.64$ & $1.96$ & $-4.82$ & $1.19$ \\
    \hline
    scalar 2SPSA & Standard & Eqs.~\eqref{eq:Gidi} & $-5.84$ & $1.08$ & $-5.84$ & $0.80$ \\
    \hline
    scalar 2CSPSA & Standard & Eqs.~\eqref{eq:Gidi} & $-5.17$ & $1.49$ & $-5.01$ & $1.08$ \\
    \hline
    QN-SPSA & Asymptotic & Eqs.~\eqref{eq:Gidi} & $-5.86$ & $0.85$ & $-6.13$ & $0.48$ \\
    \hline
    QN-CSPSA & Standard & Eqs.~\eqref{eq:Gidi} & $-5.92$ & $1.10$ & $-6.33$ & $0.53$ \\
    \hline
    scalar QN-SPSA & Asymptotic & Eqs.~\eqref{eq:Gidi} & $-5.90$ & $1.07$ & $-6.34$ & $0.53$ \\
    \hline
    scalar QN-CSPSA & Standard & Eqs.~\eqref{eq:Gidi} & $-5.93$ & $1.10$ & $-6.39$ & $0.55$ \\
    \hline
  \end{tabular}
  \caption{Best configuration and statistical indicators for each vanilla method applied to variational quantum eigensolver.}
  \label{tab:vqe_vanilla}
\end{table}


\begin{table}[ht!]
  \begin{tabular}{|c|c|c|c|c|c|c|c|c|}
    \hline
    {\bf Method} & {\bf Gains} & {\bf Post-processing} & {\bf Resampling} & {\bf Blocking} & {\bf Median} & {\bf IQR} & {\bf Mean} & {\bf STD} \\ \hline
    SPSA & Standard & - & 5 & No & $-7.00$ & $1.12$ & $-6.58$ & $0.56$ \\ \hline
    CSPSA & Standard & - & 5 & No & $-7.00$ & $1.12$  & $-6.51$ & $0.57$ \\ \hline
    2SPSA & Standard & Eqs.~~~\eqref{eq:hessian} & 2 & Yes & $-6.94$ & $1.11$ & $-6.51$ & $0.56$ \\ \hline
    2CSPSA & Standard &  Eqs.~\eqref{eq:complex_hessian} & 5 & Yes & $-6.98$ & $1.13$  & $-6.52$ & $0.56$\\ \hline
    scalar 2SPSA & Standard & Eqs.~\eqref{eq:Gidi} & 5 & Yes & $-6.97$ & $1.09$ & $-6.55$ & $0.30$ \\ \hline
    scalar 2CSPSA & Standard & Eqs.~\eqref{eq:complex_hessian} & 5 & Yes & $-6.78$ &$ 1.15$ & $-6.47$ & $0.58$ \\ \hline
    QN-SPSA & Asymptotic &  Eqs.~\eqref{eq:Gidi} & 2 & Yes & $-7.00$ & $1.12$  & $-6.50$ & $0.56$\\ \hline
    QN-CSPSA & Asymptotic & Eqs.~\eqref{eq:complex_hessian} & 5 & Yes & $-6.98$ & $1.11$ & $-6.57$ & $0.54$ \\ \hline
    scalar QN-SPSA & Asymptotic &  Eqs.~\eqref{eq:Gidi} & 5 & Yes & $-7.00$ & $1.25$ & $-6.50$ & $0.57$ \\ \hline
    scalar QN-CSPSA & Asymptotic & Eqs.~\eqref{eq:complex_hessian} & 5 & Yes & $-6.88$ & $1.13$ & $-6.51$ & $0.57$ \\ \hline
  \end{tabular}
  \caption{Best configuration and statistical indicators for each method with improvements applied to variational quantum eigensolver.}
  \label{tab:vqe_best}
\end{table}

\begin{table}[ht!]
  \begin{tabular}{|c|c|c|c|c|c|c|}\hline
    \textbf{Method} & \textbf{Gains} & \textbf{Post-processing} & \textbf{Median}             & \textbf{IQR}             & \textbf{Mean}             & \textbf{STD}                      \\ \hline
    SPSA            & Static         & -                        & $1.35\times 10^{-5} $        & $1.21\times 10^{-5} $    & $1.60\times 10^{-5} $      & $1.11\times 10^{-5}$  \\ \hline
    CSPSA           & Static         & -                        & $6.84\times 10^{-6} $        & $6.01\times 10^{-6} $    & $1.31\times 10^{-5} $      & $8.23\times 10^{-5}$  \\ \hline
    2SPSA           & Standard       & Eqs.~\eqref{eq:Gidi}     & $2.31\times 10^{-5} $        & $2.34\times 10^{-5} $    & $4.90\times 10^{-5} $      & $6.02\times 10^{-4}$  \\ \hline
    2CSPSA          & Standard       & Eqs.~\eqref{eq:Gidi}     & $8.79\times 10^{-6} $        & $8.12\times 10^{-6} $    & $1.07\times 10^{-5} $      & $9.76\times 10^{-6}$  \\ \hline
    scalar 2SPSA    & Standard       & Eqs.~\eqref{eq:Gidi}     & $1.89\times 10^{-5} $        & $1.71\times 10^{-5} $    & $4.03\times 10^{-5} $      & $4.75\times 10^{-4}$  \\ \hline
    scalar 2CSPSA   & Standard       & Eqs.~\eqref{eq:Gidi}     & $9.73\times 10^{-6} $        & $8.93\times 10^{-6} $    & $1.18\times 10^{-5} $      & $8.32\times 10^{-6}$  \\ \hline
    QN-SPSA         & Asymptotic     & Eqs.~\eqref{eq:Gidi}     & $1.93\times 10^{-5} $        & $1.88\times 10^{-5} $    & $2.36\times 10^{-5} $      & $1.82\times 10^{-5}$  \\ \hline
    QN-CSPSA        & Asymptotic     & Eqs.~\eqref{eq:Gidi}     & $9.44\times 10^{-6} $        & $8.49\times 10^{-6} $    & $1.12\times 10^{-5} $      & $7.50\times 10^{-6}$  \\ \hline
    scalar QN-SPSA  & Asymptotic     & Eqs.~\eqref{eq:Gidi}     & $3.44\times 10^{-5} $        & $3.73\times 10^{-5} $    & $5.68\times 10^{-5} $      & $1.22\times 10^{-4}$  \\ \hline
    scalar QN-CSPSA & Asymptotic     & Eqs.~\eqref{eq:Gidi}     & $1.24\times 10^{-5} $        & $1.11\times 10^{-5} $    & $1.47\times 10^{-5} $      & $1.04\times 10^{-5}$  \\ \hline
  \end{tabular}
  \caption{Best configuration and statistical indicators for each vanilla method applied to quantum control of quantum states.}
  \label{tab:qc_vanilla}
\end{table}

\begin{table}[ht!]
  \begin{tabular}{|c|c|c|c|c|c|c|c|c|}\hline
    {\textbf{Method}} & {\textbf{Gains}} & {\textbf{Post-processing}}      & {\textbf{Resampling}} & {\textbf{Blocking}} & {\textbf{Median}}            & {\textbf{IQR}}            &  {\textbf{Mean}}            & {\textbf{STD}}         \\ \hline
    {SPSA}            & {Asymptotic}     & {-}                             & {5}                   & {No}                & {$2.02\times10^{-6}$}         & {$1.75\times10^{-6}$}      & {$2.35\times10^{-6}$}       & $1.49\times10^{-6}$    \\ \hline
    {CSPSA}           & {Asymptotic}     & {-}                             & {5}                   & {No}                & {$9.24\times10^{-7}$}         & {$8.16\times10^{-7}$}      & {$1.07\times10^{-6}$}       & $6.95\times10^{-7}$    \\ \hline
    {2SPSA}           & {Standard}       & Eqs.~\eqref{eq:Gidi}            & {2}                   & {No}                & {$2.64\times10^{-5}$}         & {$2.64\times10^{-5}$}      & {$3.19\times10^{-5}$}       & $2.24\times10^{-5}$    \\ \hline
    {2CSPSA}          & {Standard}       & Eqs.~\eqref{eq:Gidi}            & {2}                   & {No}                & {$9.15\times10^{-6}$}         & {$8.45\times10^{-6}$}      & {$1.36\times10^{-5}$}       & $5.57\times10^{-5}$    \\ \hline
    {scalar 2SPSA}    & {Standard}       & Eqs.~\eqref{eq:Gidi}            & {5}                   & {No}                & {$1.43\times10^{-5}$}         & {$1.28\times10^{-5}$}      & {$1.58\times10^{-5}$}       & $9.12\times10^{-6}$    \\ \hline
    {scalar 2CSPSA}   & {Standard}       & Eqs.~\eqref{eq:Gidi}            & {5}                   & {No}                & {$8.84\times10^{-6}$}         & {$7.68\times10^{-6}$}      & {$9.85\times10^{-6}$}       & $5.84\times10^{-6}$    \\ \hline
    {QN-SPSA}         & {Asymptotic}     & Eqs.~\eqref{eq:Gidi}            & {5}                   & {Yes}               & {$3.46\times10^{-6}$}         & {$2.61\times10^{-6}$}      & {$3.85\times10^{-6}$}       & $2.22\times10^{-6}$    \\ \hline
    {QN-CSPSA}        & {Asymptotic}     & Eqs.~\eqref{eq:Gidi}            & {5}                   & {No}                & {$1.69\times10^{-6}$}         & {$1.43\times10^{-6}$}      & {$1.92\times10^{-6}$}       & $1.15\times10^{-6}$    \\ \hline
    {scalar QN-SPSA}  & {Asymptotic}     & Eqs.~\eqref{eq:Gidi}            & {5}                   & {Yes}               & {$5.92\times10^{-6}$}         & {$5.01\times10^{-6}$}      & {$6.85\times10^{-6}$}       & $4.44\times10^{-6}$    \\ \hline
    {scalar QN-CSPSA} & {Asymptotic}     & Eqs.~\eqref{eq:Gidi}            & {5}                   & {Yes}               & {$2.80\times10^{-6}$}         & {$2.24\times10^{-6}$}      & {$3.20\times10^{-6}$}       & $1.95\times10^{-6}$    \\ \hline
  \end{tabular}
  \caption{Best configuration and statistical indicators for each method with improvements applied to quantum control of quantum states.}
  \label{tab:qc_best}
\end{table}

\begin{table}[ht!]
  \begin{tabular}{|c|c|c|c|c|c|c|}
    \hline
    \textbf{Method}                  & \textbf{Gains} & \textbf{Post-processing}        & \textbf{Median} & \textbf{IQR} & \textbf{Mean} & \textbf{STD} \\ \hline
    SPSA                             & Asymptotic     & -                               &$ 4.76\times 10^{-4}$        &$ 6.48\times 10^{-5}$     &$ 4.79\times 10^{-4}$      &$ 5.50\times 10^{-5}$          \\ \hline
    CSPSA                            & Asymptotic     & -                               &$ 1.01\times 10^{-4}$        &$ 2.00\times 10^{-5}$     &$ 1.03\times 10^{-4}$      &$ 1.40\times 10^{-5}$          \\ \hline
    2SPSA                            & Standard       & Eqs.~\eqref{eq:Gidi}            &$ 6.17\times 10^{-4}$        &$ 3.52\times 10^{-3}$     &$ 3.55\times 10^{-3}$      &$ 4.75\times 10^{-4}$          \\ \hline
    2CSPSA                           & Standard       & Eqs.~\eqref{eq:Gidi}            &$ 1.43\times 10^{-4}$        &$ 8.15\times 10^{-4}$     &$ 8.15\times 10^{-4}$      &$ 1.02\times 10^{-4}$          \\ \hline
    scalar 2SPSA                     & Standard       & Eqs.~\eqref{eq:Gidi}            &$ 5.22\times 10^{-4}$        &$ 3.26\times 10^{-3}$     &$ 3.29\times 10^{-3}$      &$ 3.70\times 10^{-4}$          \\ \hline
    scalar 2CSPSA                    & Standard       & Eqs.~\eqref{eq:Gidi}            &$ 1.38\times 10^{-4}$        &$ 7.60\times 10^{-4}$     &$ 7.58\times 10^{-4}$      &$ 9.87\times 10^{-5}$          \\ \hline
    QN-SPSA                          & Standard       & Eqs.~\eqref{eq:Gidi}            &$ 6.68\times 10^{-3}$        &$ 4.04\times 10^{-3}$     &$ 6.72\times 10^{-3}$      &$ 8.42\times 10^{-4}$          \\ \hline
    QN-CSPSA                         & Standard       & Eqs.~\eqref{eq:Gidi}            &$ 1.52\times 10^{-3}$        &$ 9.43\times 10^{-4}$     &$ 1.53\times 10^{-3}$      &$ 1.93\times 10^{-4}$          \\ \hline
    scalar QN-SPSA                   & Standard       & Eqs.~\eqref{eq:Gidi}            &$ 6.58\times 10^{-3}$        &$ 4.00\times 10^{-3}$     &$ 6.55\times 10^{-3}$      &$ 8.29\times 10^{-4}$          \\ \hline
    scalar QN-CSPSA                  & Standard       & Eqs.~\eqref{eq:Gidi}            &$ 1.49\times 10^{-3}$        &$ 9.69\times 10^{-4}$     &$ 1.51\times 10^{-3}$      &$ 1.94\times 10^{-4}$          \\ \hline
  \end{tabular}
  \caption{Best configuration and statistical indicators for each vanilla method applied to self-guided quantum tomography.}
  \label{tab:sgqt_vanilla}
\end{table}

\begin{table}[ht!]
  \begin{tabular}{|c|c|c|c|c|c|c|c|c|}
    \hline
    \textbf{Method}                          & \textbf{Gains} & \textbf{Post-processing}         & \textbf{Resampling} & \textbf{Blocking} & \textbf{Median}                                & \textbf{IQR}                                 & \textbf{Mean}                                & \textbf{STD} \\ \hline
    SPSA                                     & Asymptotic     & -                                & 1                   &  No               & $4.76\times 10^{-4} $                           & $6.48\times 10^{-5} $                        & $4.79\times 10^{-4} $                         & $5.50\times 10^{-5} $                             \\ \hline
    CSPSA                                    & Asymptotic     & -                                & 1                   &  No               & $1.01\times 10^{-4} $                           & $2.00\times 10^{-5} $                        & $1.03\times 10^{-4} $                         & $1.40\times 10^{-5} $                            \\ \hline
    {2SPSA}                                  & Standard       & Eqs.~\eqref{eq:Gidi}             & 5                   &  No               & $6.45\times 10^{-4} $                           & $1.01\times 10^{-4} $                        & $6.39\times 10^{-4} $                         & $7.26\times 10^{-5} $                             \\ \hline
    {2CSPSA}                                 & Standard       & Eqs.~\eqref{eq:Gidi}             & 5                   &  No               & $1.52\times 10^{-4} $                           & $3.12\times 10^{-5} $                        & $1.54\times 10^{-4} $                         & $1.90\times 10^{-5} $                            \\ \hline
    {scalar 2SPSA}                           & Standard       & Eqs.~\eqref{eq:Gidi}             & 5                   &  No               & $1.27\times 10^{-3} $                           & $2.21\times 10^{-4} $                        & $1.27\times 10^{-3} $                         & $1.67\times 10^{-4} $                             \\ \hline
    {scalar 2CSPSA}                          & Standard       & Eqs.~\eqref{eq:Gidi}             & 5                   &  No               & $3.05\times 10^{-4} $                           & $5.05\times 10^{-5} $                        & $3.06\times 10^{-4} $                         & $4.04\times 10^{-5} $                             \\ \hline
    {QN-SPSA}                                & Standard       & Eqs.~\eqref{eq:Gidi}             & 5                   &  No               & $1.17\times 10^{-3} $                           & $2.06\times 10^{-4} $                        & $1.19\times 10^{-3} $                         & $1.45\times 10^{-4} $                             \\ \hline
    {QN-CSPSA}                               & Standard       & Eqs.~\eqref{eq:Gidi}             & 5                   &  No               & $2.74\times 10^{-4} $                           & $4.53\times 10^{-5} $                        & $2.76\times 10^{-4} $                         & $3.33\times 10^{-5} $                             \\ \hline
    {scalar QN-SPSA}                         & Standard       & Eqs.~\eqref{eq:Gidi}             & 5                   &  No               & $2.57\times 10^{-3} $                           & $4.68\times 10^{-4} $                        & $2.55\times 10^{-3} $                         & $3.45\times 10^{-4} $                             \\ \hline
    {scalar QN-CSPSA}                        & Standard       & Eqs.~\eqref{eq:Gidi}             & 5                   &  No               & $6.06\times 10^{-4} $                           & $1.15\times 10^{-4} $                        & $6.09\times 10^{-4} $                         & $7.60\times 10^{-5} $                             \\ \hline
  \end{tabular}
  \caption{Best configuration and statistical indicators for each method with improvements applied to self-guided quantum tomography.}
  \label{tab:sgqt_best}
\end{table}

\end{document}